\documentclass[sigconf]{acmart} 

\copyrightyear{2024}
\acmYear{2024}
\setcopyright{rightsretained}

\acmConference[CHI '24]{Proceedings of the CHI Conference on Human Factors in Computing Systems}{May 11--16, 2024}{Honolulu, HI, USA}
\acmBooktitle{Proceedings of the CHI Conference on Human Factors in Computing Systems (CHI '24), May 11--16, 2024, Honolulu, HI, USA}
\acmDOI{10.1145/3613904.3642382}
\acmISBN{979-8-4007-0330-0/24/05}
\begin{document}

%\title[Deepfake Pornography]{{\huge Non-Consensual Synthetic Intimate Imagery: Prevalence, Attitudes, and Knowledge in 10 Countries}}
\title[Deepfake Pornography]{Non-Consensual Synthetic Intimate Imagery: Prevalence, Attitudes, and Knowledge in 10 Countries}
\author{Rebecca Umbach}
\affiliation{\institution{Google}\country{USA}}

\author{Nicola Henry}
\affiliation{\institution{RMIT University}\country{Australia}}
\author{Gemma Beard}
\affiliation{\institution{RMIT University}\country{Australia}}
\author{Colleen Berryessa}
\affiliation{\institution{Rutgers University}\country{USA}}

\renewcommand{\shortauthors}{Umbach et al.}
%\settopmatter{printacmref=false}
%\setcopyright{none}
\renewcommand\footnotetextcopyrightpermission[1]{}

\begin{abstract}
Deepfake technologies have become ubiquitous, "democratizing" the ability to manipulate photos and videos. One popular use of deepfake technology is the creation of sexually explicit content, which can then be posted and shared widely on the internet. Drawing on a survey of over 16,000 respondents in 10 different countries, this article examines attitudes and behaviors related to "deepfake pornography" as a specific form of non-consensual synthetic intimate imagery (NSII). Our study found that deepfake pornography behaviors were considered harmful by respondents, despite nascent societal awareness. Regarding the prevalence of deepfake porn victimization and perpetration, 2.2\% of all respondents indicated personal victimization, and 1.8\% all of respondents indicated perpetration behaviors. Respondents from countries with specific legislation still reported perpetration and victimization experiences, suggesting NSII laws are inadequate to deter perpetration. Approaches to prevent and reduce harms may include digital literacy education, as well as enforced platform policies, practices, and tools which better detect, prevent, and respond to NSII content. 

\end{abstract}


\maketitle
\pagestyle{plain}
\section{Introduction}
Non-consensual synthetic intimate imagery (NSII) refers to digitally altered content that is fake but depicts the faces, bodies, and/or voices of real people. NSII can be created through more traditional means using photo-editing software to stitch together segments, add filters, or change the speed of videos --- often referred to as ``shallowfakes'' or ``cheapfakes.'' Increasingly, however, NSII is being created through the use of artificial intelligence (AI), involving different methods, such as speech-to-speech voice conversion, lip-syncing, puppet-master, face synthesis, attribute manipulation, and face-swapping \cite{wodajo2021deepfake}. AI-generated NSII is more colloquially known as ``deepfake pornography.'' The consumer creation of deepfakes (a portmanteau of ``deep learning'' and ``fake''  \cite{shahid2022matches, westerlund2019emergence}) started in late 2017 on Reddit, after a user named ``deepfakes'' posted NSII depicting the faces of female celebrities ``stitched'' onto pornographic videos \cite{kugler2021deepfake, westerlund2019emergence}. Continued consumer interest in deepfakes is reflected in the proliferation of dedicated deepfake sites and forums, often depicting celebrity targets. While deepfakes can be used in beneficial ways for accessibility and creativity \cite{farid2022creating, cover2022deepfake}, abuse potential has increased in recent years as the technology has advanced in sophistication and availability \cite{cao2023comprehensive, masood2023deepfakes, gieseke2020new, wang2022don}. Deepfakes can be weaponized and used for malicious purposes, including financial fraud, disinformation dissemination, cyberbullying, and sexual extortion (``sextortion'') \cite{farid2022creating, aliman2021transdisciplinary}. 

Non-consensual deepfake pornography can be considered a form of image-based sexual abuse because intimate images are created and/or shared without the consent of the person or persons depicted in the images. The harms of image-based sexual abuse have been well-documented, including negative impacts on victim-survivors' mental health, career prospects, and willingness to engage with others both online and offline \cite{henry2020image, citron2014criminalizing}. The proliferation of NSII technologies means that anyone can now become a victim of image-based sexual abuse, and research suggests that women continue to bear the brunt of this online abuse \cite{kugler2021deepfake, dunn2021women}. Investigating the prevalence of the misuse of deepfake technologies, specifically for NSII, can help Human Computer Interaction (HCI) research communities to better understand how best to mitigate the gendered societal harms associated with this phenomenon. Potential avenues for reduced harm may include comprehensive and considered legislation, NSII policies around the creation and/or distribution of content (e.g., Reddit's continued de-platforming of subreddits dedicated to deepfake pornography or Google's policy for removal of NSII), technical tools to help victim-survivors discover and automate takedown requests, and user experience treatments on deepfake creation tools designed to deter users from creating and distributing NSII. 

Over the past decade, the terminology and language has evolved in the image-based sexual abuse space to move beyond problematic, perpetrator terms likes ``revenge pornography'' to better encapsulate the non-consensual nature of such acts \cite{mcglynn2017beyond}. Accordingly, it is important to emphasize that deepfake pornography is, by its nature, both involuntary and non-consensual. Therefore, throughout the article, we use the broader term ``non-consensual synthetic intimate imagery'' (NSII) to refer to fake, digitally altered images created using AI and non-AI tools, as well as the more specific term ``AI-generated image-based sexual abuse'' (AI-IBSA) to refer to fake, digitally altered images created using AI. However, when reporting on our survey questions and findings, we use the more recognized ``deepfake pornography'' term (as well as ``deepfakes'') to avoid confusion for the reader.  We focus on face-swapping which involves replacing the face of the source person with the target person, so that the target person appears to engage in scenarios in which they never appeared  \cite{masood2023deepfakes}. 

To begin mapping the prevalence of and general sentiment around AI-IBSA, we addressed the following research questions: 
\begin{itemize}
    \item RQ1: What is the general public's awareness of, and attitudes towards, AI-IBSA?
    \item RQ2: What is the prevalence of AI-IBSA behaviors (e.g., creating, viewing, and or/sharing images)?
    \item RQ3: How does gender influence both RQ1 and RQ2?
 
\end{itemize} 

In this study, we surveyed over 16,000 respondents in 10 different countries. Countries were selected based on a number of considerations, including representation of the diverse approach to legislation on the issue of image-based sexual abuse, replication/expansion of previous findings (e.g., in Australia and the United States) around the prevalence of image-based sexual abuse, and geographic diversity. Questions were designed to assess attitudes towards, and experiences of, image-based sexual abuse, with specific questions relating to NSII and more specifically AI-IBSA. The study was primarily quantitative, although optional open-ended questions allowed respondents to add additional context and thoughts. The research contributes to the existing literature by: identifying trends in geographical awareness and behaviors; building on the existing image-based sexual abuse literature in terms of understanding who is most at-risk to this phenomenon; and, probing the efficacy of current or proposed remedies. 

\section{Related Work}
In this section, we provide a brief summary of the literature surrounding public perceptions of deepfakes more broadly and in relation to associated harms, deepfake detection, and legislative redress. The structure of this background review loosely follows that of our research questions; however, as prevalence data is a current gap in the literature, in our review below we also focus on whether existing or proposed laws sufficiently address the harms associated with AI-IBSA.  

\subsection{Attitudes Towards Deepfakes}

 AI-IBSA is both similar to and distinct from traditional forms of image-based sexual abuse. In introducing the ``pervert's dilemma,'' Ohman \cite{ohman2020introducing} notes that the mere creation of deepfake pornography in isolation is somewhat comparable to private sexual fantasies, which are not considered morally objectionable. However, when situated appropriately against the backdrop of gender inequality and gender-based violence, the moral unacceptability of AI-IBSA becomes clear. Moreover, deepfake technology provides a unique avenue for harassment and harm as it significantly reduces the barriers to the creation of non-consensual content \cite{akerley2020let}. As Harris \cite{harris2021video} notes, while some have argued that concerns about the epistemic effects of deepfakes are overblown (e.g., because content is unlikely to be taken at face value and consumers will be able to critically assess veracity claims), artificially generated associations can have long-term harmful effects (see also \cite{rini2022deepfakes}).

Researchers have begun to explore the different perspectives on deepfakes through attitudinal research. Although this body of work is still somewhat nascent, results suggest significant concern about the discovery and propagation of deepfake videos \cite{gottfried2019three, cochran2021deepfakes}. Additionally, studies have focused on the harms associated with deepfakes, including a study published in 2021 in which respondents deemed creating and posting deepfake pornography, even when labeled as fake, as highly harmful and deserving of punishment \cite{kugler2021deepfake}. One study \cite{fido2022celebrity} found that while female respondents indicated greater levels of victim harm, men were more likely to create deepfake pornography. The study also found that harms are greater when content is shared, as opposed to when it is used for personal sexual gratification. To the best of our knowledge, a study with Indian respondents is the sole general population survey outside of the United States. They found that awareness of deepfakes among Indian respondents was low, and even those who knew about deepfakes expected (incorrectly) to be able to easily discern fake videos \cite{shahid2022matches}. 

A subset of research has also been conducted with respondents who are knowledgeable about deepfakes. These studies have focused primarily on ethical or privacy concerns. One study examined Reddit users actively discussing deepfakes, which found, via content analysis, that users were generally pro-deepfake technology, regardless of the consequences \cite{10.1145/3491102.3517446}. In contrast, qualitative research involving individuals using an open-source deepfake creation tool found that participants had reflexively engaged with the concepts of consent and privacy, and expressed significant concerns about the misuse of deepfakes \cite{widder2022limits}. These themes were echoed and expanded upon in a mixed-methods study carried out in China \cite{li2023norms}, where users familiar with deepfakes highlighted informed consent, privacy protection, and non-deception as relevant factors influencing the social acceptance of deepfakes. That is, if deepfakes are transparent and created or distributed with the knowledge and consent of those featured, they are more acceptable than deepfakes created non-consensually and/or for deceptive purposes. In addition to these factors, attitudes are also informed by the content or aim of the deepfakes (e.g., for creative as opposed to malicious purposes). For example, two separate studies found that entertainment \cite{li2023norms} or humorous/amusement value \cite{cochran2021deepfakes} mitigated ethical judgments on deepfakes. This may suggest that content such as sexual or parody videos could be considered more acceptable than videos intended to further political disinformation, for example. 

In sum, there is some quantitative, survey-based research suggesting that average respondents have concerns about the deleterious potential of deepfakes. Similarly, existing qualitative research with participants who have expertise on deepfakes shows that they share concerns about risks and harms. More research, however, is needed to investigate geographic diversity in general attitudinal quantitative research, as well as prevalence data, to broadly understand deepfake behavior and its impacts. More work is also needed specifically on attitudes towards creating, sharing, or explicitly seeking out such content. 

\subsection{Deepfake Creation and Detection}
While researchers and journalists have been documenting the phenomenon of deepfakes for several years, interest has accelerated in recent years due to technological advancements that facilitate the creation of deepfakes in affordable and low-friction ways \cite{bond}. For example, ``deepfake pornography,'' as a keyword search within Google Scholar, returns 284 results from 2017-2019 as compared to 1,480 results from 2022-2023. Much of the research in this space \cite{gamage2021emergence} focuses on documenting how this technology works or highlighting the performance issues for various detection tools or methods \cite{10.1145/3442381.3449978, dolhansky2020deepfake, masood2023deepfakes, stroebel2023systematic, 10.1145/3579987.3586562}. 
Like other broader AI research focusing on the issues associated with biased training data sets, deepfake detection research has identified diversity and bias problems within relevant datasets \cite{xu2022comprehensive, trinh2021examination}. For example, detection techniques have been shown to be sensitive to gender, performing worse on female-based deepfakes \cite{nadimpalli2022gbdf} as compared to images of males. A recent systematic review of deepfake detection papers succinctly concludes that current models are sensitive to both novel or challenging conditions \cite{stroebel2023systematic}. This is particularly relevant, as non-digitally altered image-based sexual abuse material may be captured surreptitiously, or made with poor cameras and/or bad lighting conditions. AI-IBSA is thus likely to demonstrate the exact risk factors for poor detection rates.

In addition to technical tool failures, studies have documented the fallibility of humans in correctly identifying deepfakes as fake \cite{kobis2021fooled, korshunov2020deepfake, thaw2020real}. This is seen across a variety deepfake types, from deepfake audio \cite{mai2023warning} to photos \cite{bray2023testing, nightingale2022ai} and videos \cite{groh2022deepfake}.
It is possible that as deepfakes become more ubiquitous, humans will improve in identifying markers or other signals suggesting the content is not ``real.'' However from a technological detection perspective, it seems likely that an arms race of sorts will develop, such that increasingly sophisticated deepfakes will require significant improvements in detection methods to catch up and keep pace. 
 
\subsection{Deepfakes and the Law}
 The societal implications of deepfakes technology are far-reaching and, while many have been recognized or anticipated, experts generally agree current legislative remedies for misuse are unlikely to be effective \cite{chesney2019deep, gieseke2020new, delfino2019pornographic, mania2020legal, solichah2023protection}. Although many scholars call for legislation, there are others who are skeptical of criminalizing NSII, AI-IBSA, or deepfakes more broadly \cite{kirchengast2020deepfakes, feeney2021deepfake, widder2022limits}. Insufficient NSII laws should be considered the norm, rather than the exception, when considering the history of failed legislative proposals for image-based sexual abuse in general \cite{carter2022reflections, walker2017systematic, akerley2020let}. In the legal sphere, the first American federal law regarding deepfakes was signed in 2019. This law focuses on the potential of deepfakes to be weaponized by foreign powers to influence elections, and it also establishes a competition to encourage research into detection technologies \cite{ferraro2020federal}. Four states (California, Virginia, New York, and Georgia) have also passed laws criminalizing ``deepfake pornography'' specifically \cite{mania2020legal, ryan2022cyberflashing, kugler2021deepfake}. At the end of 2022, a federal bill was introduced in the U.S. House of Representatives specifically prohibiting deepfake pornography \cite{Morelle}. Although legislative remedies have approached political deepfakes and AI-IBSA separately, one critique from scholars is that both operate in similar ways to chill critical speech \cite{maddocks2020deepfake}. 
 
 Outside of the United States, several countries have either explicitly included NSII in existing laws or are considering making legislative amendments. For example, in Australia, some states and territories make it clear that the non-consensual distribution of intimate images includes digitally altered material as a criminal offense. The Australian Online Safety Act 2021 \cite{esafe} also provides a civil remedy scheme for a range of online harms, including the non-consensual sharing of digitally altered images. South Korea specifically prohibits NSII, alongside all other forms of pornography \cite{lee2021webcam}. And in England and Wales, the UK Online Safety Act 2023 criminalizes threatening or sharing intimate images without consent, including images that have been digitally altered \cite{Lawson, esafety}. In most countries, however, and including most of those surveyed in our study (Belgium, Denmark, France, Mexico, Netherlands, Poland, Spain), there are no protections at all.

\section{Methods}

\subsection{Ethical Considerations}
In mid-2023, we conducted a multi-country online survey on image-based sexual abuse, defined as the non-consensual taking, creating, or sharing of intimate images (photos or videos), including threats to share intimate images. The study procedure and survey were approved by the Human Research Ethics Committee at the Royal Melbourne Institute of Technology. Respondents were informed of the topic at the beginning of the survey and were allowed to opt out at any time. All questions involving self-reporting of victimization and/or perpetration were voluntary and respondents could choose an answer of ``prefer not to say'' while still completing the survey. Respondents were given support service information tailored to their geographic location at the end of the survey. Respondents were compensated based on pilot tests of time taken in each country and exchange rates.
\subsection{Study Design}
We conducted the online survey in 10 countries, employing a research and survey firm (YouGov) to survey a minimum of 1,600 adults (18+) in each country, for a total respondent count of 16,693. YouGov was selected due to their comprehensive country coverage and large panels \cite{reutersyougov}, which  reduces the likelihood of needing extensive within-country weighting. YouGov retains demographic information on their survey panelists, and screener questions were used to achieve representative sampling based on quotas. Upon qualifying for the survey, completion rates ranged from 71.0\% (Belgium) to 89\% (Australia). The final dataset consisted of respondents who passed the YouGov standard quality checks, including  bot catchers, speeder detection, and manual checking of open-text questions.  

In conducting our survey, we intentionally chose a variety of countries that represent the spectrum of how legislation has dealt with NSII, image-based abuse, and pornography more broadly. The goal was to obtain representative samples by age, gender, and location (e.g., state or territory in Australia, province in South Korea, etc.) within each country, using the latest official population estimates reflecting each surveyed country. 
The survey built on an earlier survey conducted in Australia \cite{powell2019technology}, which focused on non-consensual intimate imagery more broadly (but did not specifically ask about deepfakes). The revised survey (median completion time: 18.2 minutes) added some additional scales, teased apart how content was obtained (e.g., filmed/photographed versus stolen from a device/cloud), and included questions on NSII and ``deepfake pornography'' specifically. 

In this paper, we focus on the  subset of 23 questions in our survey relating to AI-generated image-based sexual abuse (AI-IBSA). To administer the survey in the official language(s) of each country, the survey vendor provided translations into Danish, Dutch, French, Polish, Spanish, and South Korean. These were then double-checked and edited as needed by native speaker colleagues of the research team for quality assurance. We aimed to have as similar language as possible across countries, while also accounting for cultural differences in comfort speaking about sexual topics. 

The survey was deployed from May to June 2023. In some countries, samples were not fully representative. Accordingly, when presenting disaggregated findings in the quantitative section, the data are weighted by age, gender, and location. When we aggregate across countries, the data are additionally weighted by population, rounded to the nearest 1,000. Statistics presented in the qualitative section are unweighted. 

The mean respondent age was 46.0 years (\(sd = 16.7\)). Women accounted for 50.9\% of the respondents, men for 47.6\%, and together ``other,'' ``prefer not to say,'' and ``non-binary'' accounted for the remaining 1.4\%. Detailed demographic breakdowns by country are available in Table A1 in the Appendix. Our gender results and statistical tests focus primarily on the users who identified as women or men, excluding the 1.0\% of other users for two reasons: we are using census weights, which are often limited to binary genders, and our total sample (unweighted) of non-binary respondents was 93, which limits our ability to perform statistical tests for significance. 

\subsection{Measures and Analysis}
 All quantitative analyses were performed using R Statistical Software v4.3.2 \cite{R}.  We primarily used the survey  \cite{lumley},  srvyr  \cite{ellis}, marginaleffects \cite{arel2023marginal}, and  epitools \cite{aragon} packages for statistical analyses. When calculating within-subject mean differences, we use two-tailed one-sample weighted \textit{t}-tests with bootstrapped standard errors, using 1,000 replicates (e.g., differences in criminalization attitudes across different types of behaviors). When calculating confidence intervals for proportions, we use the Rao-Scott scaled chi-squared distribution for the loglikelihood from a binomial distribution (e.g., \% of respondents who had experienced victimization). When calculating gender differences for binary categorical values, Pearson's Chi-squared tests with Rao-Scott adjustments are used to determine significance, followed by the svyglm function with a quasibinomial family to calculate risk ratios (e.g., were men more likely to report having watched deepfake pornography as compared to women, and how much more likely?).  When calculating between-gender mean differences, we use two-tailed two-sample weighted Welch \textit{t}-tests using bootstrapped standard errors, and 1,000 replicates (e.g., difference in how much women think sharing deepfake pornography of celebrities should be criminalized as compared to men).

When evaluating differences amongst categorical variables in the unweighted qualitative data, we use Pearson's Chi-squared test with Yates' continuity correction (e.g., was there a significant difference the proportion of men versus women in indicating condemnation attitudes). When examining victim-blaming attitudes by country, we use a Fisher's exact test to accommodate the small sample size. 
 
 In relation to AI-IBSA, all respondents were asked three closed-ended questions, as well as one optional open-ended question presented towards the end of the survey. They were first asked whether they were familiar with the concept of deepfake pornography using the following wording: ``Realistic-looking fake porn can now be created using artificial intelligence (AI) to swap the faces of pornographic actors with other people's faces, so that it looks like they're in a porn video. This is sometimes referred to as `deepfake pornography' as it looks very realistic and can be hard to recognize as fake or digitally created. How familiar are you with this concept?''. This question offered four possible categorical response options, as seen in Table \ref{table:familiarity}. Second, respondents were asked how much they thought a series of behaviors were worthy of criminalization (behaviors can be seen in Figure 1). Response options ranged from -2 (``Definitely should not be a crime'') to 2 (``Definitely should be a crime''), with the midpoint of 0 being ``Not sure.''  And third, respondents were asked how much they agreed or disagreed with the following statement: ``People shouldn't get upset if someone creates a digitally altered video (e.g., fake porn or a `deepfake') of them without their permission.'' Respondents answered this question on a 7-point Likert scale, anchored at 1=``Strongly disagree'' and 7=``Strongly agree'' (we present the mean alongside the median in the results - see below) \cite{sullivan2013analyzing}. 
 
 The remaining close-ended questions were asked of a subset of respondents, depending on their knowledge and experiences. A subset of respondents who indicated at least some familiarity (i.e., ``I know a little bit about it'' or ``I am quite familiar with this'') were asked follow-up questions around AI-IBSA behaviors as seen in Figure \ref{figure:experience}. 
As part of a broader set of questions regarding the victimization and perpetration of image-based sexual abuse since the age of 18, we also asked the subset of respondents who indicated a previous experience with digitally altered images whether the relevant content was a video, and if so, whether the video was a deepfake (allowing respondents to select ``Yes'', ``No'', ``Don't know'', or ``Prefer not to answer''). For victimization questions, the wording was intentionally not blame-casting. Similarly, perpetration questions were phrased neutrally to focus on whether the respondent had ever done an action. 

All respondents could opt into answering the open-ended question with the following wording: ``Would you like to add any additional thoughts on the topic of deepfakes? Please feel free to skip this question if you don't have anything to add.'' 
We analyzed the qualitative data for the open-ended response questions using an iterative, inductive approach. Our approach was to translate the responses from origin language to English via the Google Translate API, familiarize ourselves with the data, develop a codebook, code the data, and then review the data to make sense of the themes from the first coding process. To do so, two members of the research team read through 25\% of the sample responses across the countries and then co-developed a codebook that detailed descriptions and restrictions on what could be included within a code and provided concrete examples of each code. The same two members of the research team coded the first 150 responses based on four preliminary codes that emerged from the initial review of the data. Coder 1 and coder 2 independently moved through the subset of data labelling the open-ended responses with relevant codes. Following this, coders 1 and 2 compared coding allocations for quality assurance. The two coders then identified and removed responses that were deemed inappropriate or irrelevant to the research (e.g., ``no comment,'' or ``no thanks, I do not know this topic''). Finally, the team revised and discussed the themes that emerged from the responses (n = 2,474). The final codebook contained four key themes: ``condemnation,'' ``punitive and justice attitudes,'' ``a fear of technology,'' and ``victim-blaming.'' Inter-rater reliability (IRR) for each code was calculated using Cohen's Kappa. Across the four codes, Kappa values ranged from .135 to .225, with a median of .173. 

\section{Results}
Results are organized into four sections. The first section addresses the first research question, focusing on awareness and attitude questions that were asked of all respondents. The second section answers the second question by focusing on behaviors involving AI-IBSA/deepfake pornography, including victimization, perpetration, and consumption. The third section investigates the role of gender in influencing answers to the first and second research questions. Finally, the last section is a qualitative analysis of the open-ended opinion question. We present our results primarily descriptively and holistically, particularly at the country level, as we have no reason to assume differences between countries, given the lack of prior data. When discussing gender differences, statistical analyses (accounting for weights) are presented, alongside measures of uncertainty. We conclude with some qualitative trends from the single, open-end question to which respondents could choose to respond. 

\subsection{Awareness and Attitudes}
\begin{table*}[!htb]
  \captionof{table}{Respondent Familiarity with Deepfakes, by Country (weighted)}.
\centering
\label{table:familiarity}
\begin{tabular}{llrr}\hline
 \textbf{Country} & \textbf{Response}& \textbf{\% of}  &  \textbf{95\% CI}\\ &&\textbf{Respondents}\\\hline
 Australia & I have never heard of this until now & 36.8 & 35.4-38.2 \\ 
&I have only heard the phrase & 30.3 & 29-31.7 \\ 
& I know a little bit about it & 24.7 & 23.4-26 \\ 
 & I am quite familiar with this & 8.2 & 7.3-9.1 \\ 
 Belgium & I have never heard of this until now & 40.1 & 39.1-41.1 \\ 
& I have only heard the phrase & 28.0 & 27.1-28.9 \\ 
  & I know a little bit about it & 25.6 & 24.7-26.5 \\ 
 & I am quite familiar with this & 6.3 & 5.8-6.8 \\ 
 Denmark & I have never heard of this until now & 34.8 & 34.1-35.5 \\ 
& I have only heard the phrase & 44.1 & 43.4-44.8 \\ 
  & I know a little bit about it & 17.0 & 16.5-17.6 \\ 
 & I am quite familiar with this & 4.10 & 3.8-4.4 \\ 
France & I have never heard of this until now & 55.2 & 52.8-57.6 \\ 
 & I have only heard the phrase & 21.6 & 19.7-23.6 \\ 
 & I know a little bit about it & 18.4 & 16.6-20.3 \\ 
& I am quite familiar with this & 4.8 & 3.8-5.9 \\ 
Mexico & I have never heard of this until now & 51.7 & 48.8-54.7 \\ 
 & I have only heard the phrase & 22.3 & 19.9-24.8 \\ 
& I know a little bit about it & 21.0 & 18.7-23.5 \\ 
 & I am quite familiar with this & 4.9 & 3.8-6.3 \\ 
  Netherlands & I have never heard of this until now & 29.0 & 27.9-30.1 \\ 
  & I have only heard the phrase & 31.7 & 30.6-32.8 \\ 
  & I know a little bit about it & 31.8 & 30.7-32.9 \\ 
 & I am quite familiar with this & 7.5 & 6.9-8.2 \\ 
  Poland & I have never heard of this until now & 50.9 & 49.1-52.8 \\ 
& I have only heard the phrase & 30.1 & 28.4-31.8 \\ 
& I know a little bit about it & 15.2 & 13.9-16.5 \\ 
   & I am quite familiar with this & 3.8 & 3.2-4.6 \\ 
  South Korea & I have never heard of this until now & 31.1 & 29.1-33.2 \\ 
  & I have only heard the phrase & 39.2 & 37.1-41.4 \\ 
   & I know a little bit about it & 26.6 & 24.7-28.6 \\ 
    & I am quite familiar with this & 3.0 & 2.3-3.8 \\ 
  Spain & I have never heard of this until now & 49.5 & 47.5-51.5 \\ 
& I have only heard the phrase & 27.6 & 25.8-29.4 \\ 
   & I know a little bit about it & 19.5 & 17.9-21.1 \\ 
 & I am quite familiar with this & 3.5 & 2.8-4.3 \\ 
 USA & I have never heard of this until now & 42.5 & 37.7-47.4 \\ 
 & I have only heard the phrase & 26.9 & 22.7-31.4 \\ 
 & I know a little bit about it & 23.7 & 19.6-28.1 \\ 
  & I am quite familiar with this & 6.9 & 4.5-9.8 \\ 
   \hline
\end{tabular}
\end{table*}
On average, 28.1\% (\(95\% CI = 27.0\%-29.3\%\)) of respondents reported being either quite familiar with, or knowing at least a little bit about, deepfake pornography. Descriptive statistics including 95\% CIs broken out by country are available in Table \ref{table:familiarity}. The percentage of respondents indicating high familiarity ranged from 3.0\% in South Korea to 8.2\% in Australia. 

\begin{figure*}[!htb]

\includegraphics[ scale=.75] {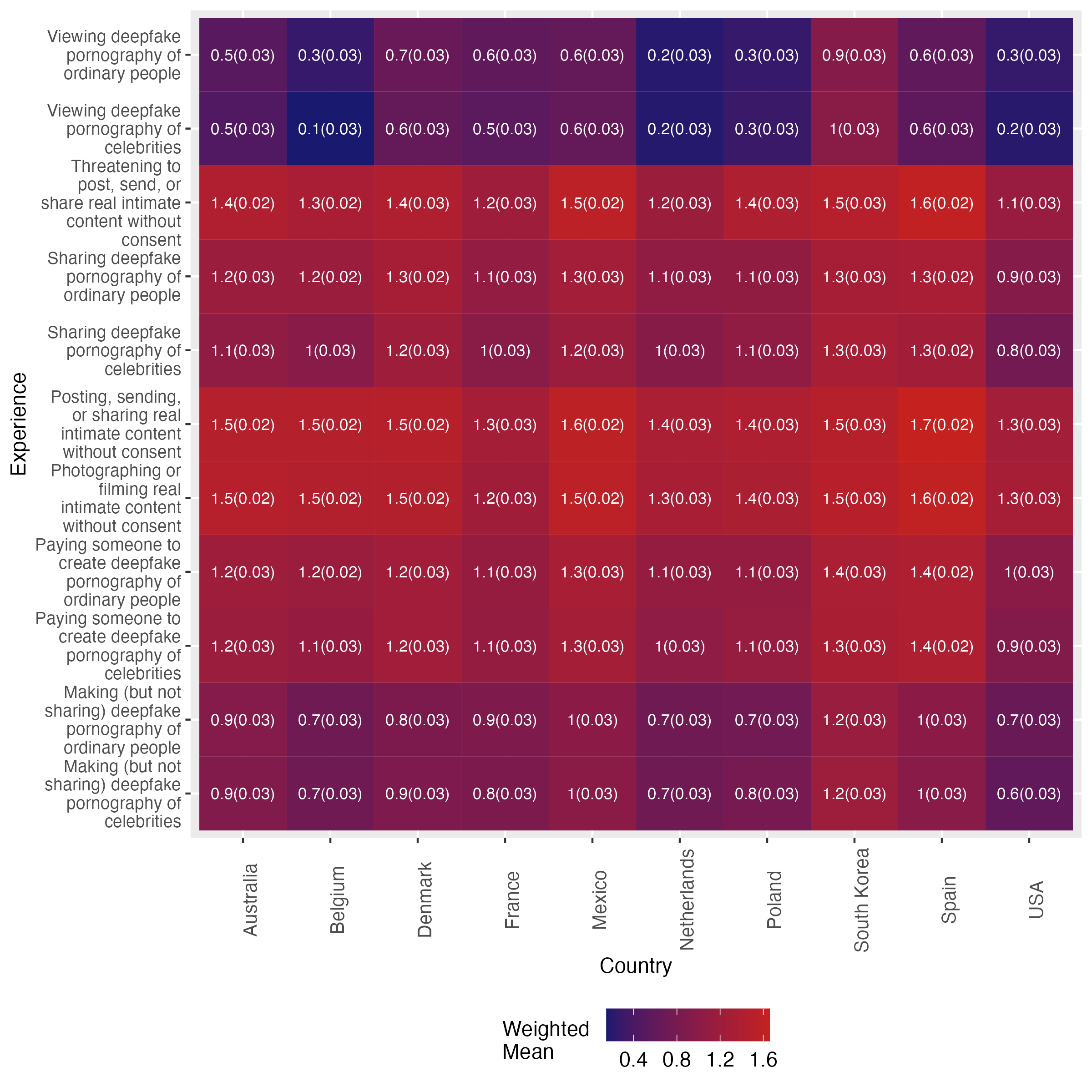}
\caption{Heatmap of Mean Criminalization Attitudes by Country. Scale ranges from -2 (``Definitely should not be a crime'') to 2 (``Definitely should be a crime''), with a midpoint of 0 ``Not sure''). Brighter red is more deserving of criminalization, darker blue is less deserving of criminalization. Standard errors are presented in parentheses.}
\label{fig:heatmapcountry}
\Description[Heat map with country on the x axis and experience type on the y axis]{Heat map with country on the x axis and experience type on the y axis. Mean weighted scores are presented alongside standard errors in parentheses. All scores are positive, indicating that on average, respondents fell on the side of ``should be a crime.''}
\end{figure*}

Because all respondents in the survey received some information about deepfake pornography through the phrasing of the awareness question, respondents were then asked whether they thought a series of actions should be a crime, with response options ranging from -2=``Definitely should not be a crime'' to 2=``Definitely should be a crime.'' The midpoint of 0 was a ``Not Sure'' option. Weighted means \cite{sullivan2013analyzing} and standard errors by country and behavior are presented in Figure \ref{fig:heatmapcountry} (with brighter red indicating more deserving of criminalization and darker blue less deserving), and weighted medians can be found in Figure A3 in the Appendix.  We intentionally included ``traditional'' forms of image-based sexual abuse (e.g., the taking, creating, threatening to share, or sharing of real non-consensual content) to better understand how users perceived AI-IBSA/deepfake pornography in comparison. Aggregated across countries, where equivalencies existed, respondents generally rated actions involving non-consensual \textit{real} content to be worse than the equivalent of AI-generated content. For example, posting/sending of real content without consent was seen as significantly more deserving of criminalization than sharing deepfake pornography of ordinary people (\(M_{difference}=0.27, t=38.76, SE=0.01, p<0.001\), Cohen's \(d\ = 0.30)\). Non-consensually filming or photographing someone was seen as significantly more deserving of criminalization as compared to making deepfake pornography (``ordinary people'' for both, \(M_{difference}=0.56, t=63.51, SE= 0.01, p<0.001\), Cohen's \(d\ = 0.50\)).

We additionally disaggregated the deepfake pornography categories by victim-type; that is, celebrities versus non-celebrities. We hypothesized that respondents would have more lenient views towards deepfake pornography behaviors involving celebrities, who are more commonly victimized than non-famous individuals. The difference between paying someone to create deepfake pornography of a celebrity versus an ordinary person was statistically significant (\(M_{difference}=0.03, t=5.01, p<0.001, SE=0.01\), Cohen's \(d\ = 0.04)\), however the effect size is negligible. Similar findings can be seen for viewing deepfake pornography of ordinary people versus celebrities (\(M_{difference}=0.03, t=4.63, p<0.001, SE=0.01\), Cohen's \(d\ = 0.03\)). This suggests little difference in perceptions of criminalization deservedness between celebrity and non-celebrity victims of equivalent behaviors.

With regard to deepfake pornography behaviors, we can draw four main conclusions. First, echoing the real content findings, the \textit{distribution} of deepfake pornography was considered especially bad, as compared to something more passive such as \textit{viewing} deepfake pornography (``ordinary people'' for both, \(M_{difference}=0.68, t=75.25, p<0.001,SE= 0.01\), Cohen's \(d\ = 0.60\)). Second, paying someone to create deepfake pornography was considered worse than making it yourself (``ordinary people'' for both, \(M_{difference}=0.33, t=46.45, p<0.001, SE= 0.01\), Cohen's \(d\ = 0.36\)). One possible interpretation is that respondents assume that going to significant lengths to have deepfake pornography created is indicative of malicious intent. Third, across countries, the behavior with the lowest average score (i.e., least bad) was the \textit{viewing} of deepfake content.

Finally, respondents were asked about how much they agreed or disagreed with the following statement: ``People shouldn't get upset if someone creates a digitally altered video (e.g., fake porn or a `deepfake') of them without their permission,'' with  anchors of 1=Strongly disagree and 7=Strongly agree. As seen in Table \ref{table:agreement}, across the different countries, the mean response fell between 1 and 2, with a median of 1 in all countries,  indicating a general sentiment that people \textit{should} get upset if they are the target of a deepfake pornography video. 

\begin{table} [!htb]
\centering
\captionof{table}{Level of agreement with statement: ``People shouldn't get upset if someone creates a digitally altered video (e.g., fake porn or a `deepfake') of them without their permission'' 
(1 = strongly disagree, 4 = neither agree not disagree, 7=strongly agree). Weighted statistics presented.} 

\begin{tabular}{lrrr}\hline
\label{table:agreement}
 \textbf{Country} & \textbf{Mean} &\textbf{SE} & \textbf{Median} \\ 
  \hline
 Australia & 1.8 & 0.04 & 1.0\\ 
 Belgium & 1.5 & 0.03 &1.0 \\ 
 Denmark & 1.4 & 0.03 & 1.0 \\ 
 France & 1.7 & 0.04 & 1.0\\ 
 Mexico & 1.5 & 0.03 & 1.0\\ 
 Netherlands & 1.5 & 0.03 &1.0\\ 
 Poland & 1.6 & 0.04 &1.0\\ 
 South Korea & 1.8 & 0.04 &1.0\\ 
 Spain & 1.4 & 0.03 &1.0\\ 
 USA & 1.8 & 0.04 &1.0\\ 
   \hline
\end{tabular}
\end{table}
\subsection{Behaviors} 
Behaviors were asked about in three different ways. First, respondents who reported that they were quite familiar with the term ``deepfakes'' or knew a little bit about it (28.1\%, \(95\% CI = 27.0\%-29.3\%\)) were asked a follow-up question about whether they had engaged in specific deepfake pornography behaviors. Second, as part of the larger survey, respondents were asked about general image-based sexual abuse victimization and perpetration. If they reported experiences with digitally altered content, they were asked follow-up questions about whether the digitally altered content involved video(s),  and if so, if the video(s) was a deepfake. This is distinct from the first question in that it a) asks for more details of the perpetration/victimization (e.g., whether they had created, threatened to share, or shared such content), and b) asks respondents who reported victimization to clarify whether they had been the victim of deepfake pornography specifically. 

\subsubsection{Behaviors of ``Familiar'' Respondents}The results of the first question (measuring perpetration behaviors) can be seen in Figure 2, where the reported data represents the percentage of all respondents. The respondents who did not see this question due to a lack of familiarity with deepfake pornography are represented in the missing data, which would make each bar sum to 100\% if included. For example, 70\% of respondents in Spain indicated they had never heard of this until now or they had only heard the phrase, so they were not asked this question. The most common behavior was the \textit{viewing} \textit{of celebrity deepfake pornography}, ranging from 2.8\% (\(95\% CI = 2.1\%-3.7\%\)) of all respondents in Poland, to 8.0\% (\(95\% CI=6.6\%-9.5\%\)) of all respondents in Australia. Aggregated across the different countries, 6.2\% (\(95\% CI=5.6\%-6.9\%\)) of all respondents indicated that they had viewed celebrity deepfake pornography. The second most common behavior (2.9\% of all respondents, \( 95\% CI = 2.5\%-3.3\%\)) was the \textit{viewing of deepfake pornography involving ``ordinary'' people}, ranging from 0.7\% (\(95\% CI = 0.4\%-1.3\%\)) in Denmark, to 5.1\%  (\(95\% CI = 4.1\%-6.2\%\)) in Mexico. Behaviors that required more time, skills, or resources were rarer, such as creating  or paying someone to create deepfake pornography of ordinary people (0.6\%, \(95\% CI=0.4\%-0.9\%\), and 0.5\%, \(95\% CI= 0.3\%-0.7\%\)), respectively, aggregated across countries)).

 \begin{figure*}[!htb]
\centering
\includegraphics[scale=1.4, width=\textwidth]{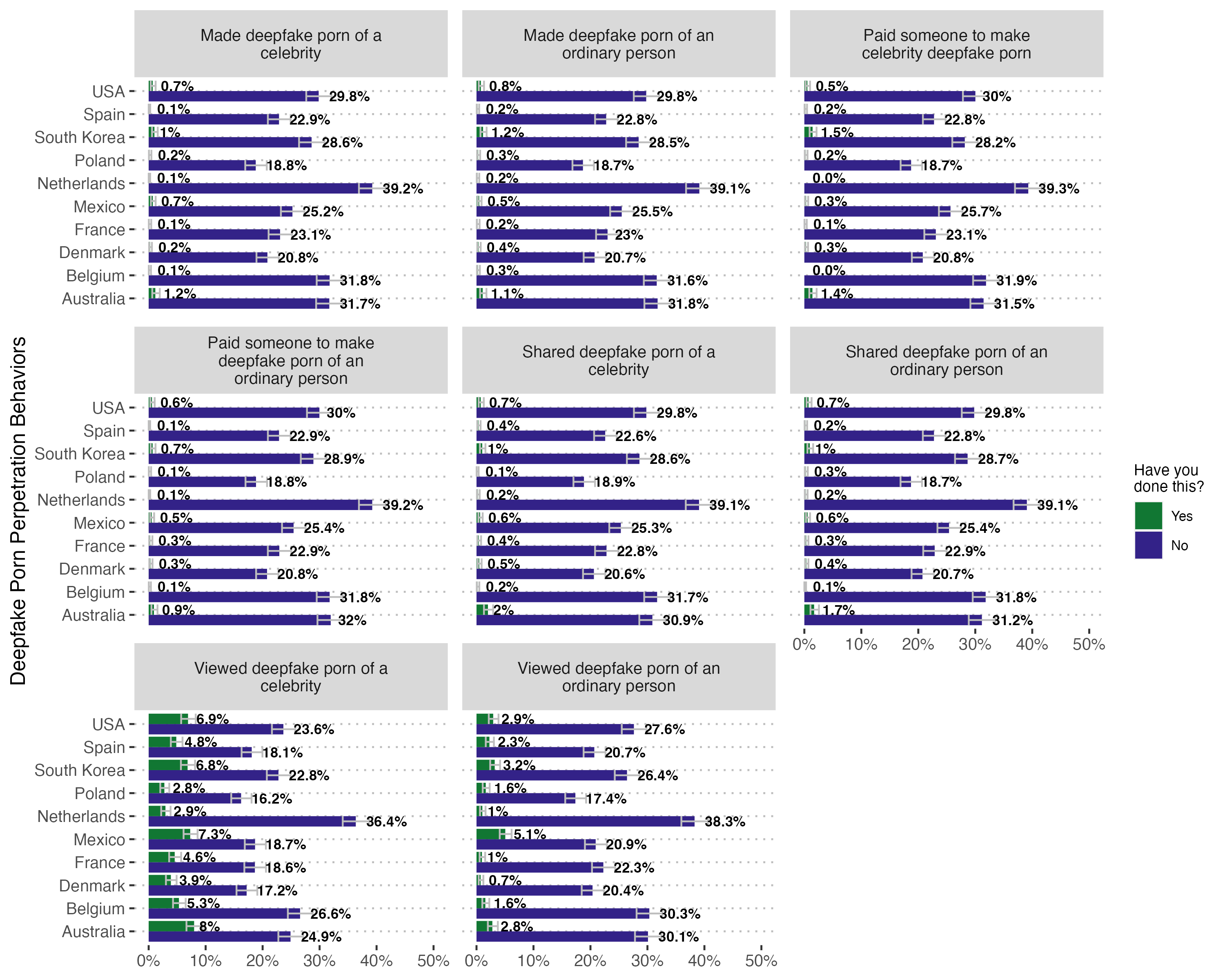}
\caption{Deepfake Pornography Behaviors by Country. Percentage of all respondents who indicated engaging, or not, in a particular behavior. 95\% confidence interval bars presented in grey.}
\label{figure:experience}
\Description{Faceted bar charts, with \% of all respondents on the x axis alongside 95\% confidence intervals, and country on the y axis. Rates are quite low for every behavior. The viewing of celebrity and non-celebrity deepfake are the most endorsed behaviors.}
\end{figure*}

\subsection{Respondents who Reported Personal Victimization or Perpetration}
\subsubsection{Victimization}
\begin{figure*}[!htb]
\centering
\includegraphics[scale=.65]{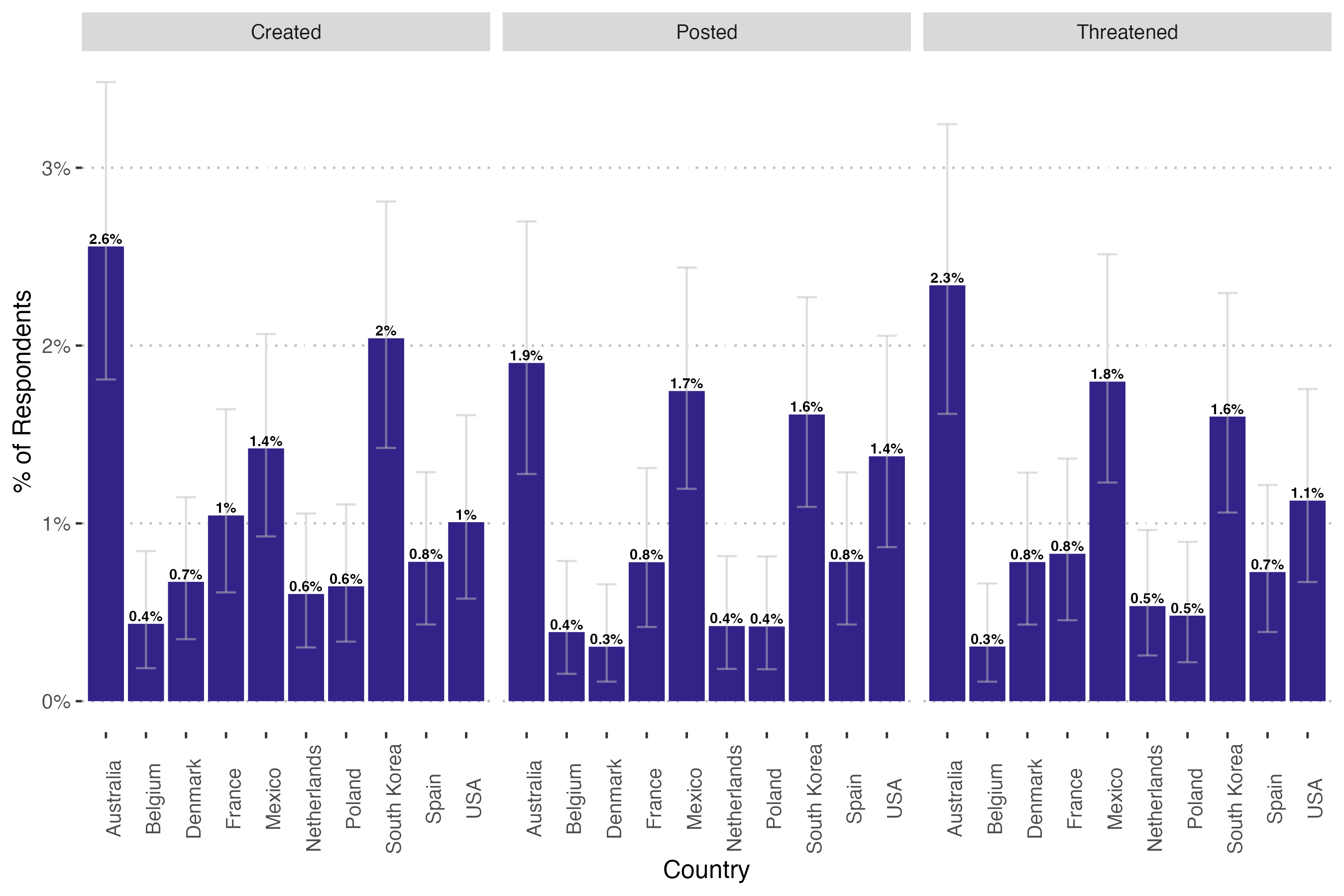}
\caption{Percentage of Respondents Reporting Different Types of Deepfake Pornography Video Victimization, by Country. 95\% CIs in grey.}
\Description{Faceted bar chart with country on the x axis and \% of respondents indicating victimization experiences on the y axis. Victimization experiences include images created, posted, sent, or shown without consent, or threats made to post, send, or show images. Rates are lowest in Belgium across behaviors, and highest in Australia, and fall between 0.3\% and 2.6\%}
\label{figure:videovic_country}
\end{figure*}

\begin{figure*}
\centering
\includegraphics[scale=.8]{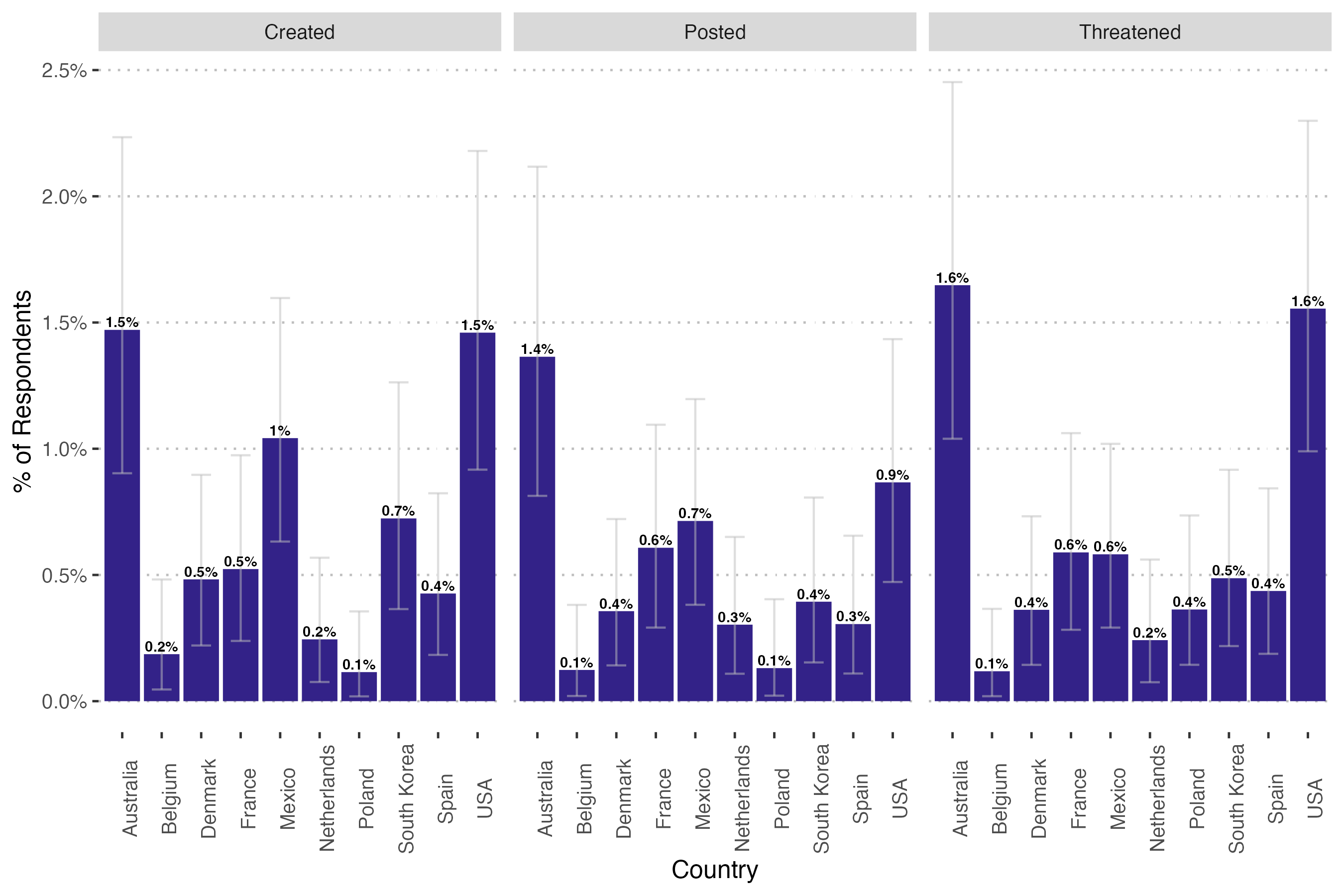}
\caption{Percentage of Respondents Reporting Different Types of Deepfake Pornography Video Perpetration, by Country. 95\% CIs in grey.}
\Description{Faceted bar chart with country on the x axis and \% of respondents indicating perpetration behaviors on the y axis. Perpetration behaviors include images created, posted, sent, or shown without consent, or threats made to post, send, or show images. Rates are low in Belgium, Poland and the Netherlands,  and highest in Australia and the United States and fall between 0.1\% and 1.6\%}
\label{figure:videoperp_country}
\end{figure*}

Experiences broken out by country are available in Figures \ref{figure:videovic_country} and \ref{figure:videoperp_country}. Aggregated across behaviors, respondents from Australia (3.7\%), Mexico (2.9\%), and South Korea (3.1\%) reported the highest rates of victimization by deepfake pornography. Full results with 95\% confidence intervals can be seen in Table A2 in the Appendix.

Aggregated across countries, 2.2\% (\(95\% CI=1.8\%-2.6\%\)) of respondents reported some form of victimization. This included: 1.2\% (\(95\% CI=0.9\%-1.5\%\)) of respondents who reported that someone had created deepfake pornography content of them; 1.3\% (\(95\% CI=1.0\%-1.6\%\)) who reported that someone had posted or sent deepfake pornography content of them; and 1.2\% (\(95\% CI=0.9\%-1.5\%\)) who reported that someone had threatened to post or send deepfake pornography content of them. 

\subsubsection{Perpetration}
Aggregating across behavior types (creating, posting, threatening to post), self-reported deepfake pornography perpetration was rare (1.8\%, \(95\% CI=1.4\%-2.2\%\)), with the highest rates reported in Australia (2.4\%) and the United States (2.6\%), and lowest in Belgium (\(0.3\%\)). Full results can be seen in Table A2 in the Appendix. Aggregating across countries, 1.0\% (\(95\% CI=0.8\%-1.4\%\)) of respondents indicated creating deepfake pornography, 1.0\% (\(95\% CI=0.8\%-1.3\%\)) reported threatening to post, send, or share deepfake pornography, and 0.7\% (\(95\% CI=0.5\%-0.9\%\)) reported actually posting, sending, or sharing deepfake pornography content. We included the creation of deepfake pornography as a perpetration behavior to mirror the fact that ``creation'' behaviors, such as filming or photographing real content without consent, fall under the umbrella of image-based sexual abuse perpetration. Nevertheless, we acknowledge that such behaviors might not be considered  ``perpetration'' by others. 

\subsection{Gender}
\subsubsection{Awareness and Attitudes}
Given some existing literature on gendered attitudes and behaviors towards image-based sexual abuse more broadly, we wanted to assess whether gender played a role in respondents' answers. There was more reported familiarity of deepfake pornography by respondents who identified as men, as opposed to those who identified as women. A Pearson's chi-squared test with Rao \& Scott adjustment confirmed the significant relationship between NSII awareness and gender (\(F = 58.7, ndf = 3.0, ddf = 49287.1, p < 0.001)\). Men were 1.92 times (\(95\% CI=1.53-2.40\)) more likely than women to report being ``quite familiar with this'' and were significantly less likely to report having never heard of deepfake pornography until now (\(RR = 0.69, 95\% CI=  0.65-0.73\)). Full details can be seen in Figure A1 of the Appendix.

Across deepfake behaviors, men rated criminalization worthiness significantly lower than women, with effect sizes falling between small and medium. For example, women found viewing deepfake pornography of celebrities (\(M_{difference}=0.58, t=29.6, SE= 0.02,\) Cohen's \(d = 0.41, p<0.001\)) and viewing deepfake pornography of non-celebrities (\(M_{difference}=0.57, t=29.2 , SE=.02,\)Cohen's \(d = 0.40, p<0.001\)) to be more deserving of criminalization as compared to men, as seen in Figure 5 (with brighter red indicating more deserving of criminalization and darker blue less deserving). The smallest difference and effect size was seen in how each gender perceived paying someone to create deepfake pornography of an ordinary person (\(M_{difference}=0.37, t=22.05 , SE=.02,\) Cohen's \(d = 0.28, p<0.001\)).
\begin{figure}[!htb]
\centering
\includegraphics[ scale=.73] {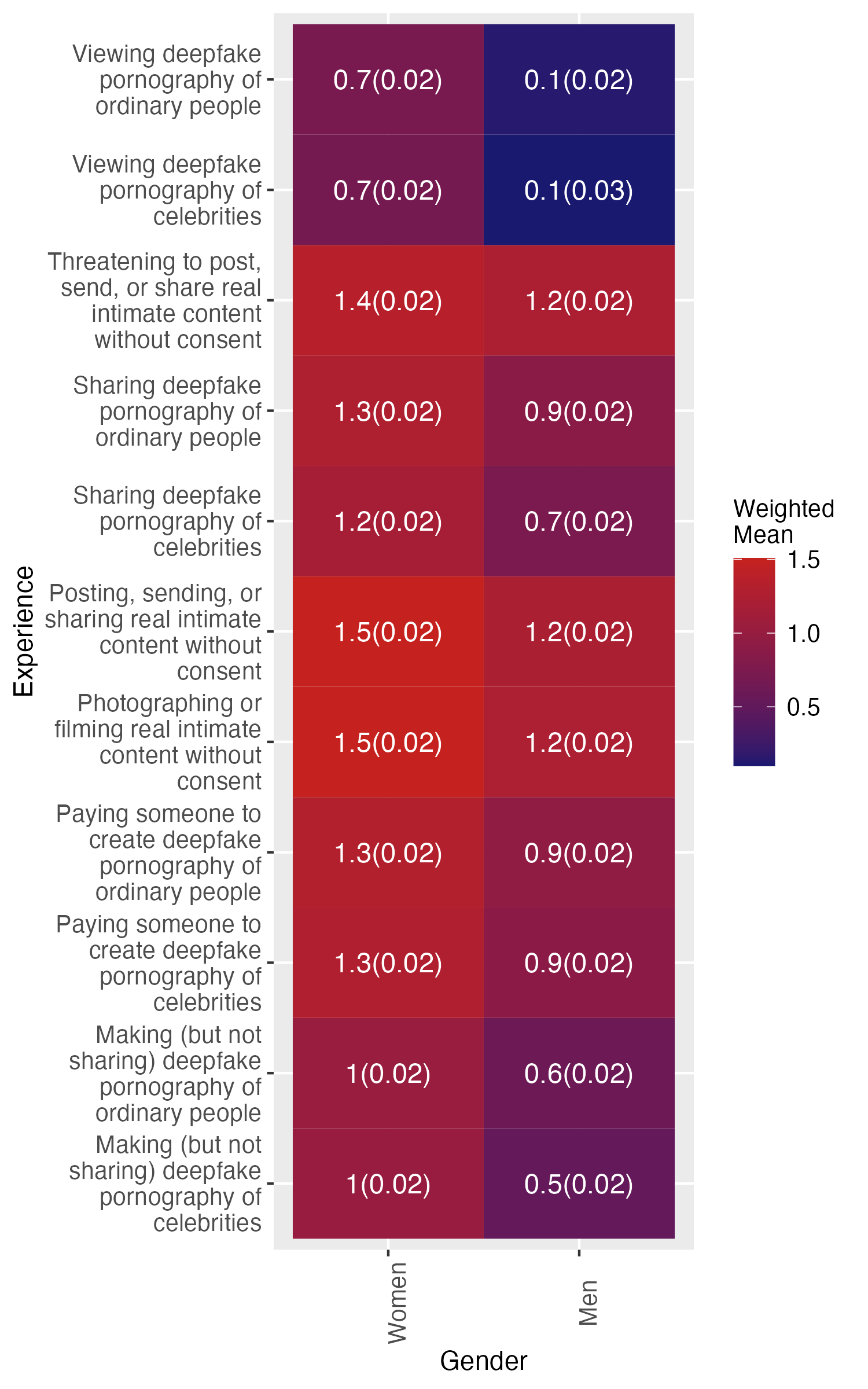}
\caption{Heatmap of Mean Criminalization Attitudes by Respondent Gender. Scale ranges from -2 (``Definitely should not be a crime'') to 2 (``Definitely should be a crime''), with a midpoint of 0 (``Not sure''). Brighter red is more deserving of criminalization, darker blue is less deserving of criminalization. Standard errors are presented in parentheses.}
\Description{Heatmap showing attitudes towards various image-based sexual abuse behaviors. The x axis is gender (women and men), and the y axis is each type of behavior (e.g., posting, sending, or showing real intimate content without consent). The behaviors involving real content all fall between ``should be a crime'' and ``definitely should be a crime,'' and all weighted means are >0, indicating that all of the behaviors fell between ``not sure'' and ``definitely should be a crime.''}
\end{figure}
 
Similarly, the mean agreement with the statement that people shouldn't be upset to be the target of deepfakes was higher in men than in women, albeit with a small effect size (\(M_{difference}=0.34, t=14.4, SE=.02\), Cohen's \(d = 0.24, p < 0.001\)).	 
 
\subsubsection{Behaviors}
Of the two most commonly reported behaviors, men respondents were more likely to report viewing deepfake pornography; both that of celebrities (\(F = 71.1, ndf = 1, ddf = 16510, p < 0.001, RR=3.5, 95\% CI=2.77-4.42\)) and non-celebrities (\(F = 14.3, ndf = 1, ddf = 16510, p < 0.001, RR=2.66, 95\% CI=1.91-3.70\)). 
\begin{figure*}[!htb]
\centering
\includegraphics[scale=.6]{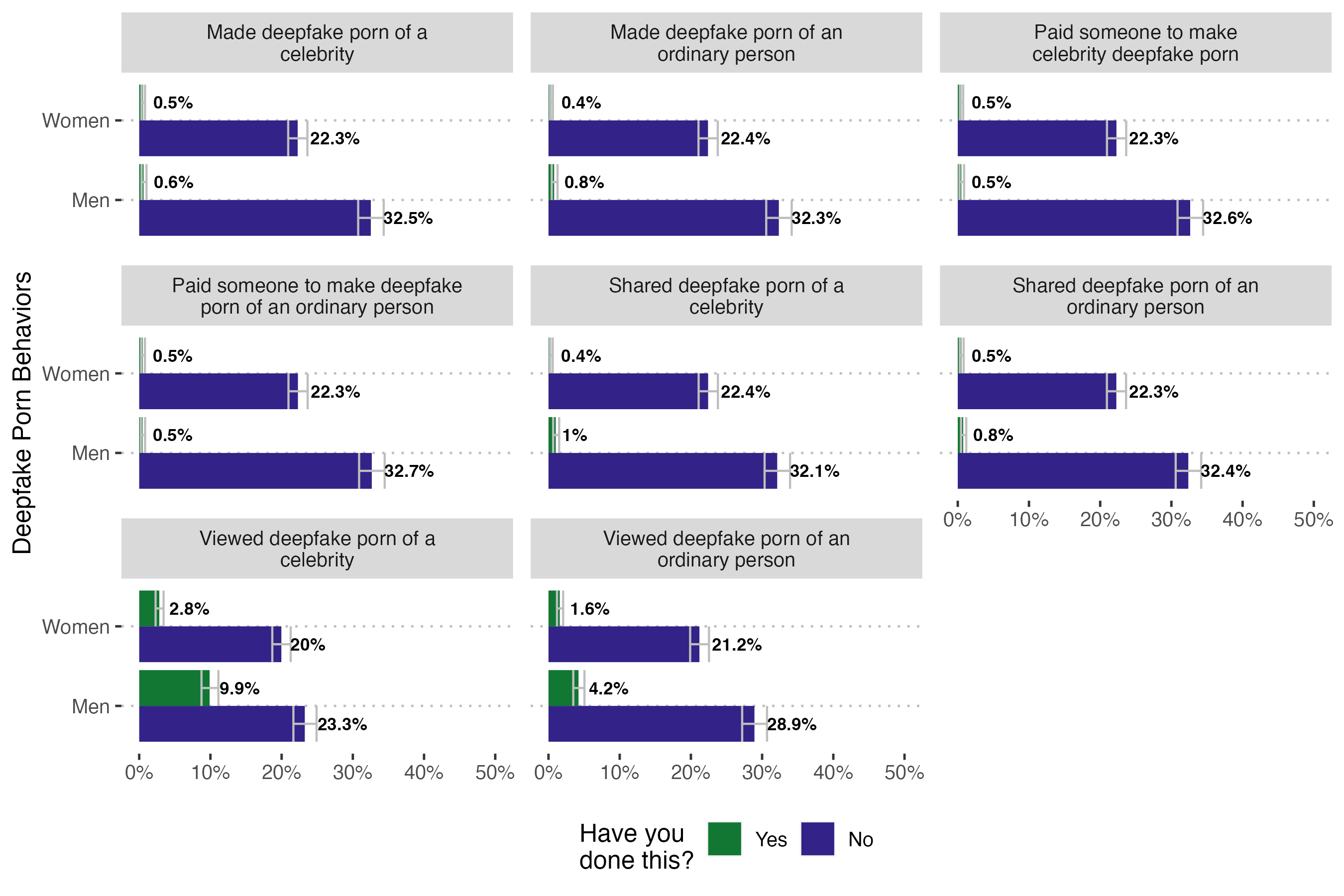}
\caption{Deepfake Pornography Behaviors by Gender. Percentage of all respondents who indicated engaging, or not, in a particular behavior. 95\% confidence interval bars presented in grey.}
\Description{Faceted bar charts, with \% of all respondents reporting they have engaged in the behavior on the x axis, and gender on the y axis. Rates were at 1\% or less for every behavior, regardless of gender, with the exception of: the viewing of celebrity deepfake pornography (9.9\% of men, 2.8\% of women), and the viewing of non-celebrity deepfake pornography (4.2\% men, 1.6\% of women).}
\label{deepfake_exp_gender}
\end{figure*}

\subsubsection{Respondents Who Reported Personal Victimization or Perpetration}
Looking at the effects of gender on victimization and perpetration, men reported significantly higher rates across all victimization and perpetration experiences related to creation and threats. For example, the risk of men reporting having been threatened with deepfake dissemination was 2.6 times that of women, and the risk of them threatening someone else was  1.9 times that of women.  Full results are presented in Table \ref{table:gender}. Notably, there were no significant differences by gender in the risk of disseminating content or having their content disseminated.

\begin{table*}[!htb]
\centering
\captionof{table}{Experiences of NSII by Gender, with Risk Ratios. } 

\begin{tabular}{llccccc}
  \hline
  \textbf{Type} &
 \textbf{Experience}   &\textbf{ Proportion of} & \textbf{Proportion of} &\textbf{\(p\)-value} &
 \textbf{RR} & \textbf{95\% CI}\\ &&\textbf{``yes'': men} & \textbf{``yes'': women}\\
  \hline
Perpetration&&&&&&\\
&Creating & 0.015 & 0.006 & <0.005 & 2.31 &1.29-4.15\\ 
& Threatening & 0.015 & 0.006 & <0.01 & 2.64 &1.40-4.98 \\ 
& Posting/Sharing & 0.009 & 0.005 & 0.16  &1.63 &0.82-3.25\\ 
Victimization &&&&&&\\
 & Created & 0.016 & 0.007 &<0.001& 2.230&1.39-3.58\\ 
&Threatened & 0.016 & 0.008 & <0.001 & 1.91 & 1.18-3.10\\ 
&Shared & 0.015 & 0.010	& 0.093 & 1.52&0.93-2.46\\ 

   \hline
   \label{table:gender}
\end{tabular}
\end{table*}

\subsection{Qualitative Responses}
In the survey, respondents were asked an optional, open-ended question: ``Would you like to add any additional thoughts on the topic of ‘deepfakes’? Please feel free to skip this question if you don’t have anything to add.'' A total of 2,477 open-ended responses were collected. Of those responses, 1,180 were removed due to inappropriate or irrelevant responses. This left us with 1,296 open-ended responses (7.8\% of respondents).  No respondents discussed their own experiences of either victimization or perpetration.

The most prevalent theme was the condemnation of deepfake pornography, with just over 43\% of the respondents who responded to this question expressing that view. Across responses, women (47.9\%) expressed greater condemnation of deepfake pornography than men (38.2\%, \(x^2 = 11.58, df = 1, p<.001\)). There were significant differences by country in terms of frequency of condemnation sentiment (\(x^2 = 42.11, df = 9, p < 0.001\)). Comments relating to the condemnation of deepfake pornography were most common from respondents from France (60.8\%), followed by Denmark (53.7\%) and the United States (48.2\%). Comments relating to this theme were less common in South Korea (25.7\%), followed by the Netherlands (27.8\%), and Belgium (35.0\%). Respondents expressed their condemnation of deepfake pornography, calling the act ``disgusting and perverted'' and ``an extremely sick concept.'' Some respondents mentioned the potential pervasive impacts of deepfake pornography, such as one Australian women (aged 35-44) who said, ``I think it’s dangerous as it could lead to serious injury (mental and emotional etc.) to the innocent person on the receiving end of the deepfake.'' Others commented on the damage it can have on a victim's employment opportunities or integrity. For example, a man from Mexico (aged 18-24) commented that ``it is a stupid strategy to continue damaging the integrity of people who do not want to be intimately seen by the public.'' There was also a general consensus among respondents that deepfakes were unethical and morally wrong. 

The second most prevalent theme, found in just over 30\% of all responses to the open-ended question, referred to a fear of technology; specifically deepfakes and artificial intelligence (AI). There were significant differences by country (\(x^2 = 32.81, df = 9, p < 0.001\)), but not by gender (\(x^2 = 0.011, df = 1, p = 0.73\)). These responses were the highest in the Netherlands (50.0\%), and the lowest in France (16.2\%) and South Korea (19.7\%). References to the futuristic nature of deepfake technology, fear of its capabilities and accessibility, as well as the realness of deepfaked images, showed up across these responses. For example, one man from Mexico (aged 25-34) commented on how ``it is difficult to assimilate [to] these types of tools, before they seemed futuristic but they are already very normal.'' Many respondents described deepfakes as a troubling and terrifying advancement of AI technology. Some, particularly women, indicated their concern about being victimized by deepfake pornography. For example, a woman from Denmark (aged 25-34)  referred to deepfakes as ``widely scary'' because ``it’s hard not to think if you have become a victim yourself without knowing it.'' Other respondents focused on the continued growth of deepfakes and the implications they might have for future generations. For example, an Australian woman (aged 45-54) described how she ``fear[ed] for [her] daughter’s future.'' Comments on the realness of deepfakes was a common thread throughout responses, with a Polish man (aged 35-44) saying ``soon it will not be known if what we [are] seeing is reality or not.'' Similarly, a Belgian woman (aged 45-54)  questioned ``can we still know if the photo or video seen or received is real or not?! It becomes difficult to make your own judgement.'' Respondents also spoke to the difficulties of regulating deepfakes online. A man from Mexico (aged 65+) emphasized how it is ``almost impossible to control,'' while a woman from the United States (aged 65+) asked ``how can one protect oneself from this? Or prove it's not them?'' 

The third most common theme was punitive and justice attitudes, seen in just over a quarter of respondents. Across responses, men were slightly more likely to express punitive and justice attitudes (\(X^2 = 4.85, df = 1, p< 0.05\)), and there were significant differences by country (\(X^2 = 91.18, df = 9, p< 0.001\)).  References to punishment, injustice, criminalization, and governmental responsibility were common in South Korea. Just over half (55.3\%) of South Korean respondents demanded strengthened laws relating to deepfakes. Responses often referred to deepfake pornography as a ``sex crime'' and called for the implementation of strong criminal justice regulations and sanctions. Accordingly, many South Korean respondents argued for the development of criminal sanctions that imposed ``strong punishment”'' for perpetrators of deepfake pornography. As one South Korean respondent (male, aged 45-54) stated, ``deepfakes [are] not yet controversial, but it can cause many social problems. In reality, legal regulations are needed to strongly punish [these] sex crimes.'' 

Many respondents disclosed that they were unaware if penalties and sanctions were available in their country for deepfakes, while others were concerned with the slow pace of governmental responses. One woman from the United States (aged 55-64) stated that she ``feared that politics doesn’t attract enough experts in tech who could help shape policy and procedure on deepfakes … and it needs to be something for which governments set as a goal to get ahead of and not ignore [or] play catch up.'' Similarly, a woman from Denmark (aged 25-34) emphasized the need to ``hurry to regulate how people are taking advantage of AI.'' References to internet regulation and platform responsibility were found across some responses. For instance, one Australian woman (aged 65+) commented that ``platforms need to take more responsibility for content,'' and a man from the Netherlands (aged 65+) recommended that this type of content should be ``immediately removed by relevant platforms and should be punishable for both maker and distributor.'' 

Finally, only about 4\% of responses indicated victim-blaming attitudes. These respondents noted that people should be more careful with what they post on the internet, and that they should refrain from taking or sharing nude photos or videos, and stop complaining about the issue and take responsibility. There were no significant differences by gender (\(X^2 = 1.68, df = 1, p=0.20\)), nor by country (per a Fisher's exact test for count data, \(p = 0.08\)).  For example, a non-binary respondent from Mexico (aged 35-44) and a woman from Poland (aged 55-64) noted that people have to be careful with what they share on the internet. Two Australian respondents encouraged people to stop taking nude images altogether; for example, one Australian man (aged 55-64) commented ``if you don’t want something shared on the internet don’t take nude photos'' while an Australian woman (aged 65+) said, ``that’s why I do NOT send photos over the internet at all. I can’t understand why people send photos to anyone --- keep your [images] private.'' One man from Poland (aged 65+) was more explicit and indicated that people should not complain about being victimized: ``you just should not share any photos on the web at all, and if you make it available, you should take into account the consequences and then regret pretending to be a fool and even worse to blame anyone.'' 

In summary, only 7.8\% of all respondents provided relevant answers to the optional, open-ended question. This means that the majority of survey respondents did not provide additional thoughts on the topic - either because they did not know enough about deepfake pornography or for other reasons. 
\section{Discussion}
Our study generated four main findings about AI-IBSA/deepfake pornography awareness, behaviors, and gender dynamics. First, despite widespread press coverage, particularly in Western countries, the concept of deepfake pornography is not well-known. Yet on balance, when informed about the concept, respondents thought behaviors associated with non-consensual fake intimate imagery should be criminalized. Second, self-reported victimization prevalence was relatively low, as compared to more traditional image-based sexual abuse prevalence rates. Third, the most common behaviors self-reported by respondents were passive and readily available (e.g., the consumption of celebrity deepfake pornography), with behaviors requiring effort and/or money being very rare (e.g., creating deepfake pornography). Finally, compared to women, men think of these behaviors as less bad, and are also more likely to report being both victimized by and perpetrating AI-IBSA. 

\subsection{Attitudes and Awareness}
Less than 30\% of all respondents indicated some level of familiarity with the concept of deepfake pornography. This was also arguably reflected in the qualitative responses, with only 7.8\% of all respondents answering the open-ended question. Presumably, as the technology becomes more widespread and ever more available, familiarity will increase. This may be a double-edged sword; it is likely to result in increased use of AI-IBSA as an abuse vector (both for purposes of humiliation, and as a tool for sexual extortion), but may also result in increased suspicion of the veracity of explicit content. 
One positive finding was general agreement across countries that perpetration behaviors associated with ``traditional'' image-based sexual abuse are deserving of criminalization.
Research has widely found some degree of victim-blaming reflected in attitudes towards traditional non-consensual intimate imagery \cite{flynn2023victim, mortreux2019understanding, sugiura2020victim}. Attitudes were slightly more lenient with regard to deepfake behaviors, although over time attitudes may evolve with increased familiarity. For instance, the realization that any private individual can be deepfaked, requiring no actions on the part of the target, may engender more empathy and less victim-blaming attitudes. It should be noted that few respondents in the qualitative part of the survey expressed victim-blaming attitudes. This can be compared to other more commonly expressed pro-social attitudes in the qualitative responses, including those expressing condemnation or support for punitive action.
\subsection{Behaviors}
The vast majority of respondents indicated no engagement in any type of AI-IBSA behaviors, with the most popular behavior being the consumption of celebrity deepfake pornography. It was even rarer for respondents to indicate perpetration themselves. This was similar for victimization, with few respondents indicating that they had been victimized by deepfake pornography. However, the victimization statistics in particular should be interpreted with caution. Even with traditional forms of image-based sexual abuse, many victims may be unaware of their victimhood. AI-IBSA may exacerbate this dynamic given that deepfake technology ``democratizes'' victimization and image-based sexual abuse by enabling anyone with a computer to be a perpetrator, and anyone with imagery online to be targeted. Also, there are additional unique motivations for AI-IBSA that may result in less likelihood of victim notification; for instance, if the motivations for the creation or distribution of AI-IBSA are for sexual gratification purposes and/or for demonstrating or practicing digital skills in AI. Finally, as content may be distributed on dedicated sites or within closed communities on sites such as Discord or Reddit, victims of AI-IBSA may not know about or find out about their images.

As a subset of image-based sexual abuse, it is useful to compare rates of AI-IBSA with previously reported statistics on image-based sexual abuse more broadly. There are several associated challenges, including sampling methodologies and definitions. For example, studies may define image-based sexual abuse behaviors differently, failing to include the non-consensual creation or filming of content, or threatening to share content, alongside the actual non-consensual sharing of content. Studies may also intentionally survey younger populations who may be more likely to have relevant experiences, as opposed to collecting population-level prevalence data. Notwithstanding those limitations, previous work has shown that image-based abuse prevalence is increasing over time, driven by changes in technology use, dating, sex, and privacy \cite{powell2020image}.  Powell et al. \cite{powell2020image} found on average, across Australia, New Zealand, and the UK, that ~1/3 of survey respondents indicated at least one experience of one or more forms of image-based sexual abuse, defined similarly to our study. An American-based study \cite{ruvalcaba2020nonconsensual} focused exclusively on the non-consensual sharing of content, and found that 1/12 respondents indicated victimization, while 1/20 reported perpetration. Even taking the more expansive definition of image-based sexual abuse, our rates of perpetration and victimization using deepfake technology were quite low. Given the trends in traditional image-based sexual abuse \cite{powell2020image}, and the increasing ease of NSII creation, we expect these rates to increase over time.  

\subsection{Gender and AI-IBSA}
Awareness of deepfakes was higher among respondents who identified as men as compared to women. This is consistent with the literature on the ``digital divide'' which suggests that women face disproportionate barriers to access and use to the internet compared to men. Even when women have access to the internet and technology, they may not have the time, resources, skills, or knowledge to leverage it \cite{acilar2023towards}. Additionally, the attitudinal questions indicate a trend towards men diminishing the harms or ``badness'' of deepfake pornography behaviors. This is consistent with previous findings on deepfake pornography \cite{fido2022celebrity}, and with other findings around image-based sexual abuse more broadly \cite{attrill2021gender, bothamley2017understanding}.

Men were more likely to report both perpetration and victimization by deepfake pornography, across subtypes. Studies continue to report mixed findings regarding the role of gender in image-based sexual abuse. Powell et al. \cite{powell2020image} reported higher rates of perpetration in men, and similar rates of victimization for men and women, but differential impacts. Our findings are somewhat inconsistent with a report which found the majority of AI-IBSA available online targeted women \cite{deeptrace}. We argue that the differences in sampling methodology may help explain these inconsistencies. For instance, in the event of sexual extortion, the content may not be made available online. It is possible that the trend seen in gender may be somewhat explained by the motivation of the perpetrator, given previous research to support higher rates of ``sextortion'' in men as compared to heterosexual women \cite{eaton2023relationship}, including for adolescent populations \cite{patchin2020sextortion}. Another explanation for this discrepancy is that research suggests most deepfake videos online (an estimated 96\%) are sexualized images of women actors and musicians, who may be less likely to participate in survey research than non-public figures like our survey respondents \cite{deeptrace}.  

While we asked broadly about motivations in the larger survey, we did not ask specifically ask about motivations for deepfake pornography perpetration or victimization. Motivations for such behaviors may range from personal sexual gratification, to use for extortion purposes, to demonstrating technical skill in the creation of content. A consideration for future research should be the use of AI-IBSA (as well as other forms of NSII not involving AI) for extortion, particularly given anecdotal evidence of significant associated harms (e.g., suicide, suicidal ideation, or suicidal attempts) \cite{10.1145/3442381.3449978}. 
\subsection{Limitations and Mitigation}
Cross-country surveys are logistically difficult and require significant consideration in interpreting findings. Our study had to address language differences and non-response bias. The former were addressed, as detailed above, through cross-validation. The latter was a more significant concern, as others have noted that language used to discuss a topic such as image-based sexual abuse and pornography more broadly will be culturally dependent \cite{schoenebeck2023online}. We attempted to mitigate this by allowing respondents to select ``Prefer not to say'' for any question that involved self-reporting on behavior, as well as periodically reminding respondents of that option throughout the survey. Nevertheless, it is possible that selection bias may have resulted in an underestimation of prevalence, with respondents opting out of the survey due to personal experiences or demographic characteristics. South Korea, for example, had a high percentage of bisexual respondents. This may indicate that our attempts at mitigating non-response bias were less successful in South Korea. In addition, in this study, we evaluated prevalence based on respondents who indicated victimization with deepfake pornography specifically. Thus, we may have underestimated prevalence because we did not include respondents who were unsure, or who incorrectly indicated that their content was not deepfaked. Finally, we note that, despite explicit goals for representative sampling, the participant pool fell short, necessitating the need to weight results. 

\subsection{Design Considerations}
In this study, we found that the majority of respondents across all countries were relatively unfamiliar with the concept of deepfake pornography. We also found that respondents, on average, found deepfake pornography behaviors to be deserving of punishment. Reported perpetration and victimization rates were low, with the most common behavior reported by respondents being the consumption of deepfake pornography. Finally, gender did play a part in both experiences and attitudes, with men deeming deepfake pornography behaviors as less harmful or deserving of punishment as compared to women. This has implications for both policy and practice.

As relatively few respondents were aware of deepfake pornography, this signals first and foremost that more education is needed to both deter potential perpetration and the consumption of AI-IBSA. Education efforts should focus on the harms suffered by victim-survivors, as well as the potential consequences of perpetration. Given the gender dynamics in terms of perceived harmfulness, prioritization around education should focus on reaching boys and men.  
\begin{figure*}[!htb]
\centering
\includegraphics[scale=.7]{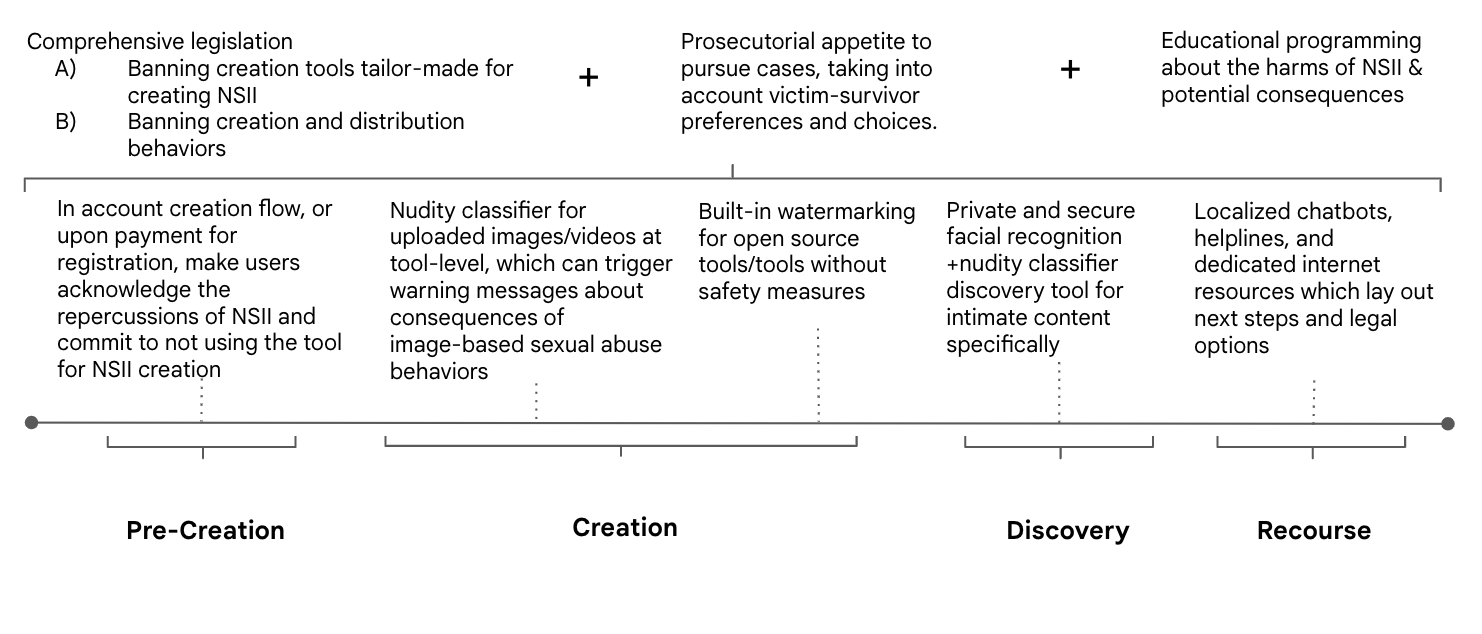}
\caption{Potential Intervention Points. The top half of this figure represents societal-level interventions, while the bottom half is more directed at individuals or companies in the deepfake econsystem.}
\label{fig:interventions}
\Description{Temporal ordering of potential intervention points, starting at the pre-creation phase (highlighting UX interventions to deter NSII content), moving into the creation phase (highlighting nudity classifiers and watermarking tools), then discovery (potential tooling around ways to discover victimization), and finally recourse (ensuring resources are up to date and useful for NSII experiences). The importance of relevant legislation and enforcement overlays all phases.}
\end{figure*}

Second, many scholars have focused on the need for legislation to criminalize these image-based abuse behaviors and provide clear redress for victims \cite{chesney2019deep, gieseke2020new, delfino2019pornographic, mania2020legal, solichah2023protection}. In order for a law to have both general and specific deterrent effects, the consequences of breaking that law has to be known by the general public, and someone who does break the law has to be prosecuted and appropriately punished. Some of the trends---particularly the high rates of prevalence in both South Korea and Australia---indicate that laws are not a panacea to eliminate or even reduce this behavior. For instance, some Australian states and territories specifically criminalize the distribution of NSII \cite{aus}, while South Korea has laws allowing for up to 3 years imprisonment for merely viewing non-consensual intimate imagery \cite{centeract,  maeng2022designing}. Given that this behavior primarily happens online, that victim-survivors may not be aware of the material, or if they become aware, may be reluctant to report it, it can be argued that existing legislation provides weak deterrence since very few people have been prosecuted for NSII in jurisdictions that do have such laws in place. As we found in this study, there was also low awareness of the concept of deepfake pornography, and one may reasonably infer low awareness of related laws. While developing legislation around an issue helps create societal norms and signals the unacceptability of a behavior, it is crucial to concede the limitations of passing regulations in enacting true societal change. The limitations of law underscores the importance of developing parallel efforts, such as technical solutions, that may mitigate these harms by reducing the creation and dissemination of NSII. 

Third, there are a number of points of potential intervention for technical solutions. First, discovery tools that allow individuals to search the web for sexual material via facial recognition technology may be empowering, although there are safety, privacy, and security concerns that need to be taken into account (e.g., building in protections such that individuals can only search for content of their own faces). Second, while many videos are posted on dedicated deepfake websites, or make no claims as to veracity, the use of deepfakes for sexual extortion indicate that technical methods of annotating content, such as watermarks, may help mitigate NSII harms \cite{harris2021video}. Third, platforms can make choices around what they allow and disallow in their policies and community standards (legislation can assist in justifying the choices made by platforms). Google Search \cite{googlepol}, Reddit \cite{verge}, and PornHub\cite{vicenews} have already published dedicated NSII policies. Fourth, various stakeholders have worked to help create solutions for victims, primarily focusing on triaging after victimization has occurred. It will be necessary for these help-seeking resources (whether helplines, online forms, or chatbots \cite{falduti2022use,  maeng2022designing}) to be aware of this new type of image-based sexual abuse and up-to-date with the types of redress that victims may seek \cite{aliman2021transdisciplinary}. Fifth, where tools do have a payment or sign up flow, users could be required as part of the flow to ingest and acknowledge information about the harms and potential consequences of NSII creation and dissemination. Finally, platforms have demonstrated an ability to detect nude content or, in some cases, non-consensual nude or sexual content. If creator tools could use those tools to block creation, or at the very least serve a warning interstitial about the harms associated with NSII (including AI-IBSA), it may act as a further deterrent for uploading this content. 

As practitioners and designers think about future considerations for intervention, it may be helpful to break this issue down into relevant behaviors as seen in Figure \ref{fig:interventions} and consider intervention options (which should be considered complementary, as opposed to solely sufficient) at each point in the process. The increased ease of access to NSII creation technologies was the catalyst for this project, and represents the first place at which interventions could be made. Are there ways, for example, to limit these tools' abilities to create nude or intimate imagery? If creators get to the point where barriers have been hacked, are there warning messages that can either highlight associated harms to targets, or potential punishments associated with the creation or distribution of such content? After creation, are there ways to build in indicators of provenance (e.g., watermarking) that signal the synthetic nature of the content? Finally, assuming most impediments can be circumvented by a motivated creator, empathetic and multi-modal resources for victims could include both tooling to discover where relevant content is published, and help for seeking take downs and recourse more broadly. These are primarily design interventions, but overarching all of these would be the societal interventions of educational programming, and relevant legislation making clear potential punishments associated with each type of behavior (extortion, posting/distribution, etc), paired with prosecutorial actions that indicate a willingness to hold perpetrators to account.

\subsection{Conclusion}
In this study, we used a survey to explore general attitudes toward AI-IBSA (or ``deepfake pornography''), as well as AI-IBSA behaviors and experiences in 10 different countries around the world. We found relatively low rates of awareness across countries, but respondents expressed concern and condemnation of image-based sexual abuse behaviors, including deepfake pornography (as seen in the questions about whether victims should get upset and criminalization, and in the qualitative responses). The most popular reported behavior was the consumption of celebrity deepfake pornography, while personal victimization and perpetration rates were quite low. Men tended to find consuming deepfake pornography less objectionable than women, which was also seen in the higher rates of self-reported consumption behavior. Finally, men were more likely to report both more perpetration and more victimization by deepfake pornography, a trend that future research should explore as we may expect prevalence to increase over time. Laws, policies, and support services can help to mitigate the harms of NSII; however, more resources need to be directed toward detecting and blocking the creation of content prior to dissemination, as well as preventing the abuse from happening in the first place through educational programs and initiatives.
\bibliographystyle{ACM-Reference-Format}
\bibliography{bib}


\begin{thebibliography}{84}

%%% the Web address.

\ifx \showCODEN    \undefined \def \showCODEN     #1{\unskip}     \fi
\ifx \showDOI      \undefined \def \showDOI       #1{#1}\fi
\ifx \showISBNx    \undefined \def \showISBNx     #1{\unskip}     \fi
\ifx \showISBNxiii \undefined \def \showISBNxiii  #1{\unskip}     \fi
\ifx \showISSN     \undefined \def \showISSN      #1{\unskip}     \fi
\ifx \showLCCN     \undefined \def \showLCCN      #1{\unskip}     \fi
\ifx \shownote     \undefined \def \shownote      #1{#1}          \fi
\ifx \showarticletitle \undefined \def \showarticletitle #1{#1}   \fi
\ifx \showURL      \undefined \def \showURL       {\relax}        \fi
% invisible to TeX
\providecommand\bibfield[2]{#2}
\providecommand\bibinfo[2]{#2}
\providecommand\natexlab[1]{#1}
\providecommand\showeprint[2][]{arXiv:#2}

\bibitem[Acilar and S{\ae}b{\o}(2023)]%
        {acilar2023towards}
\bibfield{author}{\bibinfo{person}{Ali Acilar} {and} \bibinfo{person}{{\O}ystein S{\ae}b{\o}}.} \bibinfo{year}{2023}\natexlab{}.
\newblock \showarticletitle{Towards understanding the gender digital divide: A systematic literature review}.
\newblock \bibinfo{journal}{\emph{Global Knowledge, Memory and Communication}} \bibinfo{volume}{72}, \bibinfo{number}{3} (\bibinfo{date}{Feb.} \bibinfo{year}{2023}), \bibinfo{pages}{233--249}.
\newblock
\urldef\tempurl%
\url{https://doi.org/10.3390/philosophies6010006}
\showDOI{\tempurl}


\bibitem[Adjer et~al\mbox{.}(2019)]%
        {deeptrace}
\bibfield{author}{\bibinfo{person}{Henry Adjer}, \bibinfo{person}{Giorgio Patrini}, \bibinfo{person}{Francesco Cavalli}, {and} \bibinfo{person}{Laurence Cullen}.} \bibinfo{year}{2019}\natexlab{}.
\newblock \bibinfo{booktitle}{\emph{The State of Deepfakes: Landscape, Threats, and Impact}}.
\newblock \bibinfo{type}{{T}echnical {R}eport}. \bibinfo{institution}{Deeptrace Labs}.
\newblock
\newblock
\shownote{\url{https://regmedia.co.uk/2019/10/08/deepfake_report.pdf}}.


\bibitem[Akerley(2020)]%
        {akerley2020let}
\bibfield{author}{\bibinfo{person}{Shelby Akerley}.} \bibinfo{year}{2020}\natexlab{}.
\newblock \showarticletitle{Let's talk about (fake) sex baby: A deep dive into the distributive harms of deepfake pornography}.
\newblock \bibinfo{journal}{\emph{Arizona Law Journal of Emerging Technologies}}  \bibinfo{volume}{4} (\bibinfo{year}{2020}), \bibinfo{pages}{1--58}.
\newblock


\bibitem[Aliman et~al\mbox{.}(2021)]%
        {aliman2021transdisciplinary}
\bibfield{author}{\bibinfo{person}{Nadisha-Marie Aliman}, \bibinfo{person}{Leon Kester}, {and} \bibinfo{person}{Roman Yampolskiy}.} \bibinfo{year}{2021}\natexlab{}.
\newblock \showarticletitle{Transdisciplinary AI observatory—retrospective analyses and future-oriented contradistinctions}.
\newblock \bibinfo{journal}{\emph{Philosophies}} \bibinfo{volume}{6}, \bibinfo{number}{1} (\bibinfo{date}{Jan.} \bibinfo{year}{2021}), \bibinfo{pages}{6}.
\newblock
\urldef\tempurl%
\url{https://doi.org/10.3390/philosophies6010006}
\showDOI{\tempurl}


\bibitem[Aragon et~al\mbox{.}(2020)]%
        {aragon}
\bibfield{author}{\bibinfo{person}{Tomas~J Aragon}, \bibinfo{person}{MP Fay}, \bibinfo{person}{D Wollschlaeger}, {et~al\mbox{.}}} \bibinfo{year}{2020}\natexlab{}.
\newblock \bibinfo{title}{epitools: Epidemiology Tools. R package version 0.5-10.1}.
\newblock
\newblock
\newblock
\shownote{\url{https://CRAN.R-project.org/package=epitools}}.


\bibitem[Arel-Bundock(2024)]%
        {arel2023marginal}
\bibfield{author}{\bibinfo{person}{Vincent Arel-Bundock}.} \bibinfo{year}{2024}\natexlab{}.
\newblock \bibinfo{booktitle}{\emph{marginaleffects: Predictions, Comparisons, Slopes, Marginal Means, and Hypothesis Tests}}.
\newblock
\urldef\tempurl%
\url{https://marginaleffects.com/}
\showURL{%
\tempurl}
\newblock
\shownote{R package version 0.17.0.9002}.


\bibitem[Attrill-Smith et~al\mbox{.}(2022)]%
        {attrill2021gender}
\bibfield{author}{\bibinfo{person}{Alison Attrill-Smith}, \bibinfo{person}{Caroline~J. Wesson}, \bibinfo{person}{Michelle~L. Chater}, {and} \bibinfo{person}{Lucy Weekes}.} \bibinfo{year}{2022}\natexlab{}.
\newblock \showarticletitle{Gender differences in videoed accounts of victim blaming for revenge porn for self-taken and stealth-taken sexually explicit images and videos}.
\newblock \bibinfo{journal}{\emph{Cyberpsychology: Journal of Psychosocial Research on Cyberspace}} \bibinfo{volume}{15}, \bibinfo{number}{4} (\bibinfo{date}{Apr.} \bibinfo{year}{2022}), \bibinfo{pages}{Article 3}.
\newblock
\urldef\tempurl%
\url{https://doi.org/10.5817/CP2021-4-3}
\showDOI{\tempurl}


\bibitem[Australia(2021)]%
        {aus}
\bibfield{author}{\bibinfo{person}{Youth~Law Australia}.} \bibinfo{year}{2021}\natexlab{}.
\newblock \bibinfo{title}{Image-Based Abuse | Youth Law Australia}.
\newblock
\newblock
\newblock
\shownote{\url{https://yla.org.au/nt/topics/internet-phones-and-technology/image-based-abuse/}}.


\bibitem[Bond(2023)]%
        {bond}
\bibfield{author}{\bibinfo{person}{Shannon Bond}.} \bibinfo{year}{2023}\natexlab{}.
\newblock \bibinfo{title}{It takes a few dollars and 8 minutes to create a deepfake. And that's only the start}.
\newblock
\newblock
\urldef\tempurl%
\url{https://www.npr.org/2023/03/23/1165146797/it-takes-a-few-dollars-and-8-minutes-to-create-a-deepfake-and-thats-only-the-start.}
\showURL{%
\tempurl}


\bibitem[Bothamley and Tully(2017)]%
        {bothamley2017understanding}
\bibfield{author}{\bibinfo{person}{Sarah Bothamley} {and} \bibinfo{person}{Ruth~J Tully}.} \bibinfo{year}{2017}\natexlab{}.
\newblock \showarticletitle{Understanding revenge pornography: Public perceptions of revenge pornography and victim blaming}.
\newblock \bibinfo{journal}{\emph{Journal of Aggression, Conflict and Peace Research}} \bibinfo{volume}{10}, \bibinfo{number}{1} (\bibinfo{date}{Dec.} \bibinfo{year}{2017}), \bibinfo{pages}{1--10}.
\newblock
\urldef\tempurl%
\url{https://doi.org/10.1108/JACPR-09-2016-0253}
\showDOI{\tempurl}


\bibitem[Bray et~al\mbox{.}(2023)]%
        {bray2023testing}
\bibfield{author}{\bibinfo{person}{Sergi~D Bray}, \bibinfo{person}{Shane~D Johnson}, {and} \bibinfo{person}{Bennett Kleinberg}.} \bibinfo{year}{2023}\natexlab{}.
\newblock \showarticletitle{{Testing human ability to detect "deepfake" images of human faces}}.
\newblock \bibinfo{journal}{\emph{Journal of Cybersecurity}} \bibinfo{volume}{9}, \bibinfo{number}{1} (\bibinfo{date}{June} \bibinfo{year}{2023}), \bibinfo{pages}{tyad011}.
\newblock
\showISSN{2057-2085}
\urldef\tempurl%
\url{https://doi.org/10.1093/cybsec/tyad011}
\showDOI{\tempurl}


\bibitem[Cao et~al\mbox{.}(2023)]%
        {cao2023comprehensive}
\bibfield{author}{\bibinfo{person}{Yihan Cao}, \bibinfo{person}{Siyu Li}, \bibinfo{person}{Yixin Liu}, \bibinfo{person}{Zhiling Yan}, \bibinfo{person}{Yutong Dai}, \bibinfo{person}{Philip~S. Yu}, {and} \bibinfo{person}{Lichao Sun}.} \bibinfo{year}{2023}\natexlab{}.
\newblock \bibinfo{title}{A comprehensive survey of AI-generated content (AIGC): A history of generative AI from GAN to ChatGPT}.
\newblock
\newblock
\showeprint[arxiv]{2303.04226}~[cs.AI]


\bibitem[Carter(2022)]%
        {carter2022reflections}
\bibfield{author}{\bibinfo{person}{Chance Carter}.} \bibinfo{year}{2022}\natexlab{}.
\newblock \showarticletitle{Reflections on revenge porn: Illustrating why the legal system should adopt a comprehensive response to nonconsensual pornography in the US}.
\newblock \bibinfo{journal}{\emph{Montana Law Review}}  \bibinfo{volume}{83} (\bibinfo{date}{Sept.} \bibinfo{year}{2022}), \bibinfo{pages}{293--322}.
\newblock
Issue 2.
\newblock
\shownote{\url{https://scholarworks.umt.edu/cgi/viewcontent.cgi?article=2503&context=mlr}}.


\bibitem[Center(2019)]%
        {centeract}
\bibfield{author}{\bibinfo{person}{Korea Law~Translation Center}.} \bibinfo{year}{2019}\natexlab{}.
\newblock \showarticletitle{Act On Special Cases Concerning The Punishment Of Sexual Crimes}.
\newblock  (\bibinfo{date}{Aug.} \bibinfo{year}{2019}).
\newblock
\urldef\tempurl%
\url{https://elaw.klri.re.kr/kor_service/lawView.do?hseq=51555&lang=ENG}
\showURL{%
\tempurl}


\bibitem[Chesney and Citron(2019)]%
        {chesney2019deep}
\bibfield{author}{\bibinfo{person}{Bobby Chesney} {and} \bibinfo{person}{Danielle Citron}.} \bibinfo{year}{2019}\natexlab{}.
\newblock \showarticletitle{Deep fakes: A looming challenge for privacy, democracy, and national security}.
\newblock \bibinfo{journal}{\emph{California Law Review}}  \bibinfo{volume}{107} (\bibinfo{date}{Dec.} \bibinfo{year}{2019}), \bibinfo{pages}{1753--1820}.
\newblock
\urldef\tempurl%
\url{https://doi.org/10.2139/ssrn.3213954}
\showDOI{\tempurl}


\bibitem[Citron and Franks(2014)]%
        {citron2014criminalizing}
\bibfield{author}{\bibinfo{person}{Danielle~Keats Citron} {and} \bibinfo{person}{Mary~Anne Franks}.} \bibinfo{year}{2014}\natexlab{}.
\newblock \showarticletitle{Criminalizing revenge porn}.
\newblock \bibinfo{journal}{\emph{Wake Forest Law Review}}  \bibinfo{volume}{49} (\bibinfo{date}{May} \bibinfo{year}{2014}), \bibinfo{pages}{345--391}.
\newblock


\bibitem[Cochran and Napshin(2021)]%
        {cochran2021deepfakes}
\bibfield{author}{\bibinfo{person}{Justin~D Cochran} {and} \bibinfo{person}{Stuart~A Napshin}.} \bibinfo{year}{2021}\natexlab{}.
\newblock \showarticletitle{Deepfakes: Awareness, concerns, and platform accountability}.
\newblock \bibinfo{journal}{\emph{Cyberpsychology, Behavior, and Social Networking}} \bibinfo{volume}{24}, \bibinfo{number}{3} (\bibinfo{date}{March} \bibinfo{year}{2021}), \bibinfo{pages}{164--172}.
\newblock
\urldef\tempurl%
\url{https://doi.org/10.1089/cyber.2020.0100}
\showDOI{\tempurl}


\bibitem[Cole(2018)]%
        {vicenews}
\bibfield{author}{\bibinfo{person}{Samantha Cole}.} \bibinfo{year}{2018}\natexlab{}.
\newblock \bibinfo{title}{Pornhub is banning AI-generated fake porn videos, says they're nonconsensual}.
\newblock
\newblock
\newblock
\shownote{\url{https://www.vice.com/en/article/zmwvdw/pornhub-bans-deepfakes}}.


\bibitem[Cover(2022)]%
        {cover2022deepfake}
\bibfield{author}{\bibinfo{person}{Rob Cover}.} \bibinfo{year}{2022}\natexlab{}.
\newblock \showarticletitle{Deepfake culture: The emergence of audio-video deception as an object of social anxiety and regulation}.
\newblock \bibinfo{journal}{\emph{Continuum}} \bibinfo{volume}{36}, \bibinfo{number}{4} (\bibinfo{year}{2022}), \bibinfo{pages}{609--621}.
\newblock
\urldef\tempurl%
\url{https://doi.org/10.1080/10304312.2022.2084039}
\showDOI{\tempurl}


\bibitem[Delfino(2019)]%
        {delfino2019pornographic}
\bibfield{author}{\bibinfo{person}{Rebecca~A Delfino}.} \bibinfo{year}{2019}\natexlab{}.
\newblock \showarticletitle{Pornographic deepfakes: The case for federal criminalization of revenge porn's next tragic act}.
\newblock \bibinfo{journal}{\emph{Fordham Law Review}}  \bibinfo{volume}{88} (\bibinfo{year}{2019}), \bibinfo{pages}{887--938}.
\newblock


\bibitem[Dolhansky et~al\mbox{.}(2020)]%
        {dolhansky2020deepfake}
\bibfield{author}{\bibinfo{person}{Brian Dolhansky}, \bibinfo{person}{Joanna Bitton}, \bibinfo{person}{Ben Pflaum}, \bibinfo{person}{Jikuo Lu}, \bibinfo{person}{Russ Howes}, \bibinfo{person}{Menglin Wang}, {and} \bibinfo{person}{Cristian~Canton Ferrer}.} \bibinfo{year}{2020}\natexlab{}.
\newblock \bibinfo{title}{The Deepfake Detection Challenge (DFDC) Dataset}.
\newblock
\newblock
\showeprint[arxiv]{2006.07397}~[cs.CV]


\bibitem[Dunn(2021)]%
        {dunn2021women}
\bibfield{author}{\bibinfo{person}{Suzie Dunn}.} \bibinfo{year}{2021}\natexlab{}.
\newblock \showarticletitle{Women, not politicians, are targeted most often by deepfake videos}.
\newblock \bibinfo{journal}{\emph{Centre for International Governance Innovation}} (\bibinfo{date}{March} \bibinfo{year}{2021}).
\newblock
\urldef\tempurl%
\url{https://www.cigionline.org/articles/women-not-politicians-are-targeted-most-often-deepfake-videos/?s=03.}
\showURL{%
\tempurl}


\bibitem[Eaton et~al\mbox{.}(2023)]%
        {eaton2023relationship}
\bibfield{author}{\bibinfo{person}{Asia~A Eaton}, \bibinfo{person}{Divya Ramjee}, {and} \bibinfo{person}{Jessica~F Saunders}.} \bibinfo{year}{2023}\natexlab{}.
\newblock \showarticletitle{The relationship between sextortion during COVID-19 and pre-pandemic intimate partner violence: A large study of victimization among diverse US men and women}.
\newblock \bibinfo{journal}{\emph{Victims \& Offenders}} \bibinfo{volume}{18}, \bibinfo{number}{2} (\bibinfo{year}{2023}), \bibinfo{pages}{338--355}.
\newblock
\urldef\tempurl%
\url{https://doi.org/10.1080/15564886.2021.2022057}
\showDOI{\tempurl}


\bibitem[eSafety Commissioner(2023)]%
        {esafety}
\bibfield{author}{\bibinfo{person}{eSafety Commissioner}.} \bibinfo{year}{2023}\natexlab{}.
\newblock \bibinfo{title}{Image-Based Abuse}.
\newblock
\newblock
\urldef\tempurl%
\url{https://www.esafety.gov.au/key-topics/image-based-abuse.}
\showURL{%
\tempurl}


\bibitem[Falduti and Tessaris(2022)]%
        {falduti2022use}
\bibfield{author}{\bibinfo{person}{Mattia Falduti} {and} \bibinfo{person}{Sergio Tessaris}.} \bibinfo{year}{2022}\natexlab{}.
\newblock \showarticletitle{On the use of chatbots to report non-consensual intimate images abuses: The legal expert perspective}. In \bibinfo{booktitle}{\emph{Proceedings of the 2022 ACM Conference on Information Technology for Social Good}}. \bibinfo{pages}{96--102}.
\newblock


\bibitem[Farid(2022)]%
        {farid2022creating}
\bibfield{author}{\bibinfo{person}{Hany Farid}.} \bibinfo{year}{2022}\natexlab{}.
\newblock \showarticletitle{Creating, using, misusing, and detecting deep fakes}.
\newblock \bibinfo{journal}{\emph{Journal of Online Trust and Safety}} \bibinfo{volume}{1}, \bibinfo{number}{4} (\bibinfo{date}{Sept.} \bibinfo{year}{2022}).
\newblock
\urldef\tempurl%
\url{https://doi.org/10.54501/jots.v1i4.56}
\showDOI{\tempurl}


\bibitem[Feeney(2021)]%
        {feeney2021deepfake}
\bibfield{author}{\bibinfo{person}{Matthew Feeney}.} \bibinfo{year}{2021}\natexlab{}.
\newblock \showarticletitle{Deepfake laws risk creating more problems than they solve}.
\newblock \bibinfo{journal}{\emph{Regulatory Transparency Project}} (\bibinfo{year}{2021}).
\newblock
\urldef\tempurl%
\url{https://rtp.fedsoc.org/wp-content/uploads/Paper-Deepfake-Laws-Risk-Creating-More-Problems-Than-They-Solve.pdf.}
\showURL{%
\tempurl}


\bibitem[Ferraro et~al\mbox{.}(2020)]%
        {ferraro2020federal}
\bibfield{author}{\bibinfo{person}{Matthew~F Ferraro}, \bibinfo{person}{Jason~C Chipman}, {and} \bibinfo{person}{Stephen~W Preston}.} \bibinfo{year}{2020}\natexlab{}.
\newblock \showarticletitle{The federal "deepfakes" law}.
\newblock \bibinfo{journal}{\emph{The Journal of Robotics, Artificial Intelligence \& Law}} \bibinfo{volume}{3}, \bibinfo{number}{4} (\bibinfo{year}{2020}), \bibinfo{pages}{229--233}.
\newblock


\bibitem[Fido et~al\mbox{.}(2022)]%
        {fido2022celebrity}
\bibfield{author}{\bibinfo{person}{Dean Fido}, \bibinfo{person}{Jaya Rao}, {and} \bibinfo{person}{Craig~A Harper}.} \bibinfo{year}{2022}\natexlab{}.
\newblock \showarticletitle{Celebrity status, sex, and variation in psychopathy predicts judgements of and proclivity to generate and distribute deepfake pornography}.
\newblock \bibinfo{journal}{\emph{Computers in Human Behavior}}  \bibinfo{volume}{129} (\bibinfo{date}{April} \bibinfo{year}{2022}), \bibinfo{pages}{107141}.
\newblock
\urldef\tempurl%
\url{https://doi.org/10.1016/j.chb.2021.107141}
\showDOI{\tempurl}


\bibitem[Flynn et~al\mbox{.}(2023)]%
        {flynn2023victim}
\bibfield{author}{\bibinfo{person}{Asher Flynn}, \bibinfo{person}{Elena Cama}, \bibinfo{person}{Anastasia Powell}, {and} \bibinfo{person}{Adrian~J Scott}.} \bibinfo{year}{2023}\natexlab{}.
\newblock \showarticletitle{Victim-blaming and image-based sexual abuse}.
\newblock \bibinfo{journal}{\emph{Journal of Criminology}} \bibinfo{volume}{56}, \bibinfo{number}{1} (\bibinfo{year}{2023}), \bibinfo{pages}{7--25}.
\newblock
\urldef\tempurl%
\url{https://doi.org/10.1177/263380762211353}
\showDOI{\tempurl}


\bibitem[{Freedman Ellis} and Schneider(2023)]%
        {ellis}
\bibfield{author}{\bibinfo{person}{Greg {Freedman Ellis}} {and} \bibinfo{person}{Ben Schneider}.} \bibinfo{year}{2023}\natexlab{}.
\newblock \bibinfo{booktitle}{\emph{srvyr: 'dplyr'-Like Syntax for Summary Statistics of Survey Data}}.
\newblock
\urldef\tempurl%
\url{https://CRAN.R-project.org/package=srvyr}
\showURL{%
\tempurl}
\newblock
\shownote{R package version 1.2.0}.


\bibitem[Gamage et~al\mbox{.}(2022)]%
        {10.1145/3491102.3517446}
\bibfield{author}{\bibinfo{person}{Dilrukshi Gamage}, \bibinfo{person}{Piyush Ghasiya}, \bibinfo{person}{Vamshi Bonagiri}, \bibinfo{person}{Mark~E. Whiting}, {and} \bibinfo{person}{Kazutoshi Sasahara}.} \bibinfo{year}{2022}\natexlab{}.
\newblock \showarticletitle{Are deepfakes concerning? Analyzing conversations of deepfakes on Reddit and exploring societal implications}. In \bibinfo{booktitle}{\emph{Proceedings of the 2022 CHI Conference on Human Factors in Computing Systems}} (New Orleans, LA, USA) \emph{(\bibinfo{series}{CHI '22})}. \bibinfo{publisher}{Association for Computing Machinery}, \bibinfo{address}{New York, NY, USA}, Article \bibinfo{articleno}{103}, \bibinfo{numpages}{19}~pages.
\newblock
\showISBNx{9781450391573}
\urldef\tempurl%
\url{https://doi.org/10.1145/3491102.3517446}
\showDOI{\tempurl}


\bibitem[Gamage et~al\mbox{.}(2021)]%
        {gamage2021emergence}
\bibfield{author}{\bibinfo{person}{Dilrukshi Gamage}, \bibinfo{person}{Kazutoshi Sasahara}, {and} \bibinfo{person}{Jiayu Chen}.} \bibinfo{year}{2021}\natexlab{}.
\newblock \showarticletitle{The emergence of deepfakes and its societal implications: A systematic review.}
\newblock \bibinfo{journal}{\emph{TTO}} (\bibinfo{date}{Oct.} \bibinfo{year}{2021}), \bibinfo{pages}{28--39}.
\newblock


\bibitem[Gieseke(2020)]%
        {gieseke2020new}
\bibfield{author}{\bibinfo{person}{Anne~Pechenik Gieseke}.} \bibinfo{year}{2020}\natexlab{}.
\newblock \showarticletitle{"The new weapon of choice": Law's current inability to properly address deepfake pornography}.
\newblock \bibinfo{journal}{\emph{Vanderbilt Law Review}}  \bibinfo{volume}{73} (\bibinfo{year}{2020}), \bibinfo{pages}{1479--1515}.
\newblock


\bibitem[Google(2023)]%
        {googlepol}
\bibfield{author}{\bibinfo{person}{Google}.} \bibinfo{year}{2023}\natexlab{}.
\newblock \bibinfo{title}{Remove involuntary fake pornography from Google}.
\newblock
\newblock
\urldef\tempurl%
\url{https://support.google.com/websearch/answer/9116649?hl=en.}
\showURL{%
\tempurl}


\bibitem[Gottfried(2019)]%
        {gottfried2019three}
\bibfield{author}{\bibinfo{person}{Jeffrey Gottfried}.} \bibinfo{year}{2019}\natexlab{}.
\newblock \bibinfo{title}{About three-quarters of Americans favor steps to restrict altered videos and images}.
\newblock
\newblock
\urldef\tempurl%
\url{https://www.pewresearch.org/short-reads/2019/06/14/about-three-quarters-of-americans-favor-steps-to-restrict-altered-videos-and-images/.}
\showURL{%
\tempurl}


\bibitem[Groh et~al\mbox{.}(2022)]%
        {groh2022deepfake}
\bibfield{author}{\bibinfo{person}{Matthew Groh}, \bibinfo{person}{Ziv Epstein}, \bibinfo{person}{Chaz Firestone}, {and} \bibinfo{person}{Rosalind Picard}.} \bibinfo{year}{2022}\natexlab{}.
\newblock \showarticletitle{Deepfake detection by human crowds, machines, and machine-informed crowds}.
\newblock \bibinfo{journal}{\emph{Proceedings of the National Academy of Sciences}} \bibinfo{volume}{119}, \bibinfo{number}{1} (\bibinfo{year}{2022}), \bibinfo{pages}{e2110013119}.
\newblock


\bibitem[Harris(2021)]%
        {harris2021video}
\bibfield{author}{\bibinfo{person}{Keith~Raymond Harris}.} \bibinfo{year}{2021}\natexlab{}.
\newblock \showarticletitle{Video on demand: What deepfakes do and how they harm}.
\newblock \bibinfo{journal}{\emph{Synthese}} \bibinfo{volume}{199}, \bibinfo{number}{5-6} (\bibinfo{year}{2021}), \bibinfo{pages}{13373--13391}.
\newblock
\urldef\tempurl%
\url{https://doi.org/10.1007/s11229-021-03379-y}
\showDOI{\tempurl}


\bibitem[Henry et~al\mbox{.}(2020)]%
        {henry2020image}
\bibfield{author}{\bibinfo{person}{Nicola Henry}, \bibinfo{person}{Clare McGlynn}, \bibinfo{person}{Asher Flynn}, \bibinfo{person}{Kelly Johnson}, \bibinfo{person}{Anastasia Powell}, {and} \bibinfo{person}{Adrian~J Scott}.} \bibinfo{year}{2020}\natexlab{}.
\newblock \bibinfo{booktitle}{\emph{Image-based sexual abuse: A study on the causes and consequences of non-consensual nude or sexual imagery}}.
\newblock \bibinfo{publisher}{Routledge, New York, NY}.
\newblock
\showISBNx{9780367524401}


\bibitem[Hollister(2023)]%
        {verge}
\bibfield{author}{\bibinfo{person}{Sean Hollister}.} \bibinfo{year}{2023}\natexlab{}.
\newblock \bibinfo{title}{Reddit's AI porn ban has a carveout for Rule 34}.
\newblock
\newblock
\urldef\tempurl%
\url{https://www.theverge.com/2023/7/6/21525998/reddit-ai-porn-fictional-characters-carveout-sexual-images.}
\showURL{%
\tempurl}


\bibitem[Kirchengast(2020)]%
        {kirchengast2020deepfakes}
\bibfield{author}{\bibinfo{person}{Tyrone Kirchengast}.} \bibinfo{year}{2020}\natexlab{}.
\newblock \showarticletitle{Deepfakes and image manipulation: criminalisation and control}.
\newblock \bibinfo{journal}{\emph{Information \& Communications Technology Law}} \bibinfo{volume}{29}, \bibinfo{number}{3} (\bibinfo{year}{2020}), \bibinfo{pages}{308--323}.
\newblock
\urldef\tempurl%
\url{https://doi.org/10.1080/13600834.2020.1794615}
\showDOI{\tempurl}


\bibitem[K{\"o}bis et~al\mbox{.}(2021)]%
        {kobis2021fooled}
\bibfield{author}{\bibinfo{person}{Nils~C K{\"o}bis}, \bibinfo{person}{Barbora Dole{\v{z}}alov{\'a}}, {and} \bibinfo{person}{Ivan Soraperra}.} \bibinfo{year}{2021}\natexlab{}.
\newblock \showarticletitle{Fooled twice: People cannot detect deepfakes but think they can}.
\newblock \bibinfo{journal}{\emph{Iscience}} \bibinfo{volume}{24}, \bibinfo{number}{11} (\bibinfo{year}{2021}).
\newblock
\urldef\tempurl%
\url{https://doi.org/10.1016/j.isci.2021.103364}
\showDOI{\tempurl}


\bibitem[Korshunov and Marcel(2020)]%
        {korshunov2020deepfake}
\bibfield{author}{\bibinfo{person}{Pavel Korshunov} {and} \bibinfo{person}{Sébastien Marcel}.} \bibinfo{year}{2020}\natexlab{}.
\newblock \bibinfo{title}{Deepfake detection: Humans vs. machines}.
\newblock
\newblock
\showeprint[arxiv]{2009.03155}~[cs.CV]


\bibitem[Kugler and Pace(2021)]%
        {kugler2021deepfake}
\bibfield{author}{\bibinfo{person}{Matthew~B Kugler} {and} \bibinfo{person}{Carly Pace}.} \bibinfo{year}{2021}\natexlab{}.
\newblock \showarticletitle{Deepfake privacy: Attitudes and regulation}.
\newblock \bibinfo{journal}{\emph{Northwestern University Law Review}}  \bibinfo{volume}{116} (\bibinfo{year}{2021}), \bibinfo{pages}{611--680}.
\newblock
\urldef\tempurl%
\url{https://doi.org/ssrn.3781968}
\showDOI{\tempurl}


\bibitem[Lawson(2023)]%
        {Lawson}
\bibfield{author}{\bibinfo{person}{Amanda Lawson}.} \bibinfo{year}{2023}\natexlab{}.
\newblock \showarticletitle{A look at global deepfake regulation approaches}.
\newblock \bibinfo{journal}{\emph{Responsible Artificial Intelligence Institute}} (\bibinfo{date}{April} \bibinfo{year}{2023}).
\newblock
\urldef\tempurl%
\url{https://www.responsible.ai/post/a-look-at-global-deepfake-regulation-approaches.}
\showURL{%
\tempurl}


\bibitem[Lee(2021)]%
        {lee2021webcam}
\bibfield{author}{\bibinfo{person}{Min~Joo Lee}.} \bibinfo{year}{2021}\natexlab{}.
\newblock \showarticletitle{Webcam modelling in Korea: Censorship, pornography, and eroticism}.
\newblock \bibinfo{journal}{\emph{Porn Studies}} \bibinfo{volume}{8}, \bibinfo{number}{4} (\bibinfo{date}{June} \bibinfo{year}{2021}), \bibinfo{pages}{485--498}.
\newblock
\urldef\tempurl%
\url{https://doi.org/10.1080/23268743.2021.1901602}
\showDOI{\tempurl}


\bibitem[Li and Wan(2023)]%
        {li2023norms}
\bibfield{author}{\bibinfo{person}{Minghui Li} {and} \bibinfo{person}{Yan Wan}.} \bibinfo{year}{2023}\natexlab{}.
\newblock \showarticletitle{Norms or fun? The influence of ethical concerns and perceived enjoyment on the regulation of deepfake information}.
\newblock \bibinfo{journal}{\emph{Internet Research}} (\bibinfo{year}{2023}).
\newblock


\bibitem[Lumley(2023)]%
        {lumley}
\bibfield{author}{\bibinfo{person}{Thomas Lumley}.} \bibinfo{year}{2023}\natexlab{}.
\newblock \bibinfo{title}{Survey: Analysis of Complex Survey Samples}.
\newblock
\newblock
\newblock
\shownote{R package version 4.2}.


\bibitem[Maddocks(2020)]%
        {maddocks2020deepfake}
\bibfield{author}{\bibinfo{person}{Sophie Maddocks}.} \bibinfo{year}{2020}\natexlab{}.
\newblock \showarticletitle{"A deepfake porn plot intended to silence me": Exploring continuities between pornographic and "political" deep fakes}.
\newblock \bibinfo{journal}{\emph{Porn Studies}} \bibinfo{volume}{7}, \bibinfo{number}{4} (\bibinfo{year}{2020}), \bibinfo{pages}{415--423}.
\newblock
\urldef\tempurl%
\url{https://doi.org/10.1080/23268743.2020.1757499}
\showDOI{\tempurl}


\bibitem[Maeng and Lee(2022)]%
        {maeng2022designing}
\bibfield{author}{\bibinfo{person}{Wookjae Maeng} {and} \bibinfo{person}{Joonhwan Lee}.} \bibinfo{year}{2022}\natexlab{}.
\newblock \showarticletitle{Designing and evaluating a chatbot for survivors of image-based sexual abuse}. In \bibinfo{booktitle}{\emph{Proceedings of the 2022 CHI Conference on Human Factors in Computing Systems}}. \bibinfo{pages}{1--21}.
\newblock


\bibitem[Mai et~al\mbox{.}(2023)]%
        {mai2023warning}
\bibfield{author}{\bibinfo{person}{Kimberly~T Mai}, \bibinfo{person}{Sergi~D Bray}, \bibinfo{person}{Toby Davies}, {and} \bibinfo{person}{Lewis~D Griffin}.} \bibinfo{year}{2023}\natexlab{}.
\newblock \showarticletitle{Warning: Humans cannot reliably detect speech deepfakes}.
\newblock \bibinfo{journal}{\emph{arXiv preprint arXiv:2301.07829}} (\bibinfo{year}{2023}).
\newblock


\bibitem[Mania(2020)]%
        {mania2020legal}
\bibfield{author}{\bibinfo{person}{Karolina Mania}.} \bibinfo{year}{2020}\natexlab{}.
\newblock \showarticletitle{The legal implications and remedies concerning revenge porn and fake porn: A common law perspective}.
\newblock \bibinfo{journal}{\emph{Sexuality \& Culture}} \bibinfo{volume}{24}, \bibinfo{number}{6} (\bibinfo{year}{2020}), \bibinfo{pages}{2079--2097}.
\newblock


\bibitem[Masood et~al\mbox{.}(2023)]%
        {masood2023deepfakes}
\bibfield{author}{\bibinfo{person}{Momina Masood}, \bibinfo{person}{Mariam Nawaz}, \bibinfo{person}{Khalid~Mahmood Malik}, \bibinfo{person}{Ali Javed}, \bibinfo{person}{Aun Irtaza}, {and} \bibinfo{person}{Hafiz Malik}.} \bibinfo{year}{2023}\natexlab{}.
\newblock \showarticletitle{Deepfakes generation and detection: State-of-the-art, open challenges, countermeasures, and way forward}.
\newblock \bibinfo{journal}{\emph{Applied Intelligence}} \bibinfo{volume}{53}, \bibinfo{number}{4} (\bibinfo{year}{2023}), \bibinfo{pages}{3974--4026}.
\newblock


\bibitem[McGlynn et~al\mbox{.}(2017)]%
        {mcglynn2017beyond}
\bibfield{author}{\bibinfo{person}{Clare McGlynn}, \bibinfo{person}{Erika Rackley}, {and} \bibinfo{person}{Ruth Houghton}.} \bibinfo{year}{2017}\natexlab{}.
\newblock \showarticletitle{Beyond "revenge porn": The continuum of image-based sexual abuse}.
\newblock \bibinfo{journal}{\emph{Feminist Legal Studies}}  \bibinfo{volume}{25} (\bibinfo{year}{2017}), \bibinfo{pages}{25--46}.
\newblock
\urldef\tempurl%
\url{https://doi.org/https//:10.1007/s10691-017-9343-2}
\showDOI{\tempurl}


\bibitem[Morelle(2023)]%
        {Morelle}
\bibfield{author}{\bibinfo{person}{Joseph~D. Morelle}.} \bibinfo{year}{2023}\natexlab{}.
\newblock \bibinfo{title}{Preventing Deepfakes of Intimate Images Act}.
\newblock
\newblock
\urldef\tempurl%
\url{https://www.congress.gov/bill/117th-congress/house-bill/9631/text.}
\showURL{%
\tempurl}


\bibitem[Mortreux et~al\mbox{.}(2019)]%
        {mortreux2019understanding}
\bibfield{author}{\bibinfo{person}{Colette Mortreux}, \bibinfo{person}{Karen Kellard}, \bibinfo{person}{Nicola Henry}, {and} \bibinfo{person}{Asher Flynn}.} \bibinfo{year}{2019}\natexlab{}.
\newblock \showarticletitle{Understanding the attitudes and motivations of adults who engage in image-based abuse}.
\newblock  (\bibinfo{year}{2019}).
\newblock
\urldef\tempurl%
\url{https://www.esafety.gov.au/sites/default/files/2019-10/Research_Report_IBA_Perp_Motivations.pdf.}
\showURL{%
\tempurl}


\bibitem[Nadimpalli and Rattani(2022)]%
        {nadimpalli2022gbdf}
\bibfield{author}{\bibinfo{person}{Aakash~Varma Nadimpalli} {and} \bibinfo{person}{Ajita Rattani}.} \bibinfo{year}{2022}\natexlab{}.
\newblock \showarticletitle{GBDF: Gender balanced deepfake dataset towards fair deepfake detection}.
\newblock \bibinfo{journal}{\emph{arXiv preprint arXiv:2207.10246}} (\bibinfo{year}{2022}).
\newblock


\bibitem[Nightingale and Farid(2022)]%
        {nightingale2022ai}
\bibfield{author}{\bibinfo{person}{Sophie~J Nightingale} {and} \bibinfo{person}{Hany Farid}.} \bibinfo{year}{2022}\natexlab{}.
\newblock \showarticletitle{AI-synthesized faces are indistinguishable from real faces and more trustworthy}.
\newblock \bibinfo{journal}{\emph{Proceedings of the National Academy of Sciences}} \bibinfo{volume}{119}, \bibinfo{number}{8} (\bibinfo{year}{2022}), \bibinfo{pages}{e2120481119}.
\newblock


\bibitem[of~the~eSafety Commissioner~Australia(2023)]%
        {esafe}
\bibfield{author}{\bibinfo{person}{Office of~the~eSafety Commissioner~Australia}.} \bibinfo{year}{2023}\natexlab{}.
\newblock \bibinfo{title}{Regulatory Schemes}.
\newblock
\newblock
\newblock
\shownote{\url{https://www.esafety.gov.au/about-us/who-we-are/regulatory-schemes##image-based-abuse-scheme}}.


\bibitem[{\"O}hman(2020)]%
        {ohman2020introducing}
\bibfield{author}{\bibinfo{person}{Carl {\"O}hman}.} \bibinfo{year}{2020}\natexlab{}.
\newblock \showarticletitle{Introducing the pervert’s dilemma: A contribution to the critique of deepfake pornography}.
\newblock \bibinfo{journal}{\emph{Ethics and Information Technology}} \bibinfo{volume}{22}, \bibinfo{number}{2} (\bibinfo{year}{2020}), \bibinfo{pages}{133--140}.
\newblock
\urldef\tempurl%
\url{https://doi.org/10.1007/s10676-019-09522-1}
\showDOI{\tempurl}


\bibitem[Patchin and Hinduja(2020)]%
        {patchin2020sextortion}
\bibfield{author}{\bibinfo{person}{Justin~W Patchin} {and} \bibinfo{person}{Sameer Hinduja}.} \bibinfo{year}{2020}\natexlab{}.
\newblock \showarticletitle{Sextortion among adolescents: Results from a national survey of US youth}.
\newblock \bibinfo{journal}{\emph{Sexual Abuse}} \bibinfo{volume}{32}, \bibinfo{number}{1} (\bibinfo{year}{2020}), \bibinfo{pages}{30--54}.
\newblock
\urldef\tempurl%
\url{https://doi.org/10.1177/1079063218800469}
\showDOI{\tempurl}


\bibitem[Powell and Henry(2019)]%
        {powell2019technology}
\bibfield{author}{\bibinfo{person}{Anastasia Powell} {and} \bibinfo{person}{Nicola Henry}.} \bibinfo{year}{2019}\natexlab{}.
\newblock \showarticletitle{Technology-facilitated sexual violence victimization: Results from an online survey of Australian adults}.
\newblock \bibinfo{journal}{\emph{Journal of Interpersonal Violence}} \bibinfo{volume}{34}, \bibinfo{number}{17} (\bibinfo{year}{2019}), \bibinfo{pages}{3637--3665}.
\newblock
\urldef\tempurl%
\url{https://doi.org/10.1177/0886260516672}
\showDOI{\tempurl}


\bibitem[Powell et~al\mbox{.}(2020)]%
        {powell2020image}
\bibfield{author}{\bibinfo{person}{Anastasia Powell}, \bibinfo{person}{Adrian Scott}, \bibinfo{person}{Asher Flynn}, {and} \bibinfo{person}{Nicola Henry}.} \bibinfo{year}{2020}\natexlab{}.
\newblock \showarticletitle{Image-based sexual abuse: An international study of victims and perpetrators – A summary report}.
\newblock  (\bibinfo{year}{2020}).
\newblock
\urldef\tempurl%
\url{https://research.monash.edu/en/publications/image-based-sexual-abuse-an-international-study-of-victims-and-pe.}
\showURL{%
\tempurl}


\bibitem[Pu et~al\mbox{.}(2021)]%
        {10.1145/3442381.3449978}
\bibfield{author}{\bibinfo{person}{Jiameng Pu}, \bibinfo{person}{Neal Mangaokar}, \bibinfo{person}{Lauren Kelly}, \bibinfo{person}{Parantapa Bhattacharya}, \bibinfo{person}{Kavya Sundaram}, \bibinfo{person}{Mobin Javed}, \bibinfo{person}{Bolun Wang}, {and} \bibinfo{person}{Bimal Viswanath}.} \bibinfo{year}{2021}\natexlab{}.
\newblock \showarticletitle{Deepfake videos in the wild: Analysis and detection}. In \bibinfo{booktitle}{\emph{Proceedings of the Web Conference 2021}} (Ljubljana, Slovenia) \emph{(\bibinfo{series}{WWW '21})}. \bibinfo{publisher}{Association for Computing Machinery}, \bibinfo{address}{New York, NY, USA}, \bibinfo{pages}{981–992}.
\newblock
\showISBNx{9781450383127}
\urldef\tempurl%
\url{https://doi.org/10.1145/3442381.3449978}
\showDOI{\tempurl}


\bibitem[{R Core Team}(2023)]%
        {R}
\bibfield{author}{\bibinfo{person}{{R Core Team}}.} \bibinfo{year}{2023}\natexlab{}.
\newblock \bibinfo{booktitle}{\emph{R: A Language and Environment for Statistical Computing}}.
\newblock R Foundation for Statistical Computing, Vienna, Austria.
\newblock
\urldef\tempurl%
\url{https://www.R-project.org/}
\showURL{%
\tempurl}


\bibitem[Rana and Sung(2023)]%
        {10.1145/3579987.3586562}
\bibfield{author}{\bibinfo{person}{Md~Shohel Rana} {and} \bibinfo{person}{Andrew~H. Sung}.} \bibinfo{year}{2023}\natexlab{}.
\newblock \showarticletitle{Deepfake detection: A tutorial}. In \bibinfo{booktitle}{\emph{Proceedings of the 9th ACM International Workshop on Security and Privacy Analytics}} (Charlotte, NC, USA) \emph{(\bibinfo{series}{IWSPA '23})}. \bibinfo{publisher}{Association for Computing Machinery}, \bibinfo{address}{New York, NY, USA}, \bibinfo{pages}{55–56}.
\newblock
\showISBNx{9798400700996}
\urldef\tempurl%
\url{https://doi.org/10.1145/3579987.3586562}
\showDOI{\tempurl}


\bibitem[Rini and Cohen(2022)]%
        {rini2022deepfakes}
\bibfield{author}{\bibinfo{person}{Regina Rini} {and} \bibinfo{person}{Leah Cohen}.} \bibinfo{year}{2022}\natexlab{}.
\newblock \showarticletitle{Deepfakes, deep harms}.
\newblock \bibinfo{journal}{\emph{Journal of Ethics \& Social Philosophy}}  \bibinfo{volume}{22} (\bibinfo{year}{2022}).
\newblock
Issue 2.
\urldef\tempurl%
\url{https://doi.org/10.26556/jesp.v22i2.1628}
\showDOI{\tempurl}


\bibitem[Ruvalcaba and Eaton(2020)]%
        {ruvalcaba2020nonconsensual}
\bibfield{author}{\bibinfo{person}{Yanet Ruvalcaba} {and} \bibinfo{person}{Asia~A Eaton}.} \bibinfo{year}{2020}\natexlab{}.
\newblock \showarticletitle{Nonconsensual pornography among US adults: A sexual scripts framework on victimization, perpetration, and health correlates for women and men.}
\newblock \bibinfo{journal}{\emph{Psychology of Violence}} \bibinfo{volume}{10}, \bibinfo{number}{1} (\bibinfo{year}{2020}), \bibinfo{pages}{68--78}.
\newblock
\urldef\tempurl%
\url{https://doi.org/10.1037/vio0000233}
\showDOI{\tempurl}


\bibitem[Ryan-White(2022)]%
        {ryan2022cyberflashing}
\bibfield{author}{\bibinfo{person}{Georgina Ryan-White}.} \bibinfo{year}{2022}\natexlab{}.
\newblock \showarticletitle{Cyberflashing and deepfake pornography}.
\newblock  (\bibinfo{year}{2022}).
\newblock
\urldef\tempurl%
\url{http://www.niassembly.gov.uk/globalassets/documents/raise/publications/2017-2022/2022/justice/0122.pdf}
\showURL{%
\tempurl}


\bibitem[Sandle(2021)]%
        {reutersyougov}
\bibfield{author}{\bibinfo{person}{Paul Sandle}.} \bibinfo{year}{2021}\natexlab{}.
\newblock \showarticletitle{UK's YouGov says demand from Silicon Valley clients holding up}.
\newblock \bibinfo{journal}{\emph{Reuters}} (\bibinfo{date}{March} \bibinfo{year}{2021}).
\newblock
\urldef\tempurl%
\url{https://www.reuters.com/business/uks-yougov-says-demand-silicon-valley-clients-holding-up-2023-03-21/.}
\showURL{%
\tempurl}


\bibitem[Schoenebeck et~al\mbox{.}(2023)]%
        {schoenebeck2023online}
\bibfield{author}{\bibinfo{person}{Sarita Schoenebeck}, \bibinfo{person}{Amna Batool}, \bibinfo{person}{Giang Do}, \bibinfo{person}{Sylvia Darling}, \bibinfo{person}{Gabriel Grill}, \bibinfo{person}{Daricia Wilkinson}, \bibinfo{person}{Mehtab Khan}, \bibinfo{person}{Kentaro Toyama}, {and} \bibinfo{person}{Louise Ashwell}.} \bibinfo{year}{2023}\natexlab{}.
\newblock \showarticletitle{Online harassment in majority contexts: Examining harms and remedies across countries}. In \bibinfo{booktitle}{\emph{Proceedings of the 2023 CHI Conference on Human Factors in Computing Systems}}. \bibinfo{pages}{1--16}.
\newblock


\bibitem[Shahid et~al\mbox{.}(2022)]%
        {shahid2022matches}
\bibfield{author}{\bibinfo{person}{Farhana Shahid}, \bibinfo{person}{Srujana Kamath}, \bibinfo{person}{Annie Sidotam}, \bibinfo{person}{Vivian Jiang}, \bibinfo{person}{Alexa Batino}, {and} \bibinfo{person}{Aditya Vashistha}.} \bibinfo{year}{2022}\natexlab{}.
\newblock \showarticletitle{”It matches my worldview”: Examining perceptions and attitudes around fake videos}. In \bibinfo{booktitle}{\emph{Proceedings of the 2022 CHI Conference on Human Factors in Computing Systems}}. \bibinfo{pages}{1--15}.
\newblock


\bibitem[Solichah et~al\mbox{.}(2023)]%
        {solichah2023protection}
\bibfield{author}{\bibinfo{person}{Isnaini~Imroatus Solichah}, \bibinfo{person}{Faizin Sulistio}, {and} \bibinfo{person}{Milda Istiqomah}.} \bibinfo{year}{2023}\natexlab{}.
\newblock \showarticletitle{Protection of victims of deep fake pornography in a legal perspective in Indonesia}.
\newblock \bibinfo{journal}{\emph{International Journal of Multicultural and Multireligious Understanding}} \bibinfo{volume}{10}, \bibinfo{number}{1} (\bibinfo{year}{2023}), \bibinfo{pages}{383--390}.
\newblock


\bibitem[Stroebel et~al\mbox{.}(2023)]%
        {stroebel2023systematic}
\bibfield{author}{\bibinfo{person}{Laura Stroebel}, \bibinfo{person}{Mark Llewellyn}, \bibinfo{person}{Tricia Hartley}, \bibinfo{person}{Tsui~Shan Ip}, {and} \bibinfo{person}{Mohiuddin Ahmed}.} \bibinfo{year}{2023}\natexlab{}.
\newblock \showarticletitle{A systematic literature review on the effectiveness of deepfake detection techniques}.
\newblock \bibinfo{journal}{\emph{Journal of Cyber Security Technology}} \bibinfo{volume}{7}, \bibinfo{number}{2} (\bibinfo{year}{2023}), \bibinfo{pages}{83--113}.
\newblock
\urldef\tempurl%
\url{https://doi.org/10.1080/23742917.2023.2192888}
\showURL{%
\tempurl}


\bibitem[Sugiura and Smith(2020)]%
        {sugiura2020victim}
\bibfield{author}{\bibinfo{person}{Lisa Sugiura} {and} \bibinfo{person}{April Smith}.} \bibinfo{year}{2020}\natexlab{}.
\newblock \showarticletitle{Victim blaming, responsibilization and resilience in online sexual abuse and harassment}.
\newblock \bibinfo{journal}{\emph{Victimology: Research, Policy and Activism}} (\bibinfo{year}{2020}), \bibinfo{pages}{45--79}.
\newblock


\bibitem[Sullivan and Artino~Jr(2013)]%
        {sullivan2013analyzing}
\bibfield{author}{\bibinfo{person}{Gail~M Sullivan} {and} \bibinfo{person}{Anthony~R Artino~Jr}.} \bibinfo{year}{2013}\natexlab{}.
\newblock \showarticletitle{Analyzing and interpreting data from Likert-type scales}.
\newblock \bibinfo{journal}{\emph{Journal of Graduate Medical Education}} \bibinfo{volume}{5}, \bibinfo{number}{4} (\bibinfo{year}{2013}), \bibinfo{pages}{541--542}.
\newblock
\urldef\tempurl%
\url{https://doi.org/JGME-5-4-18}
\showDOI{\tempurl}


\bibitem[Thaw et~al\mbox{.}(2020)]%
        {thaw2020real}
\bibfield{author}{\bibinfo{person}{Nyein~Nyein Thaw}, \bibinfo{person}{Thin July}, \bibinfo{person}{Aye~Nu Wai}, \bibinfo{person}{Dion Hoe-Lian Goh}, {and} \bibinfo{person}{Alton~YK Chua}.} \bibinfo{year}{2020}\natexlab{}.
\newblock \showarticletitle{Is it real? A study on detecting deepfake videos}.
\newblock \bibinfo{journal}{\emph{Proceedings of the Association for Information Science and Technology}} \bibinfo{volume}{57}, \bibinfo{number}{1} (\bibinfo{year}{2020}), \bibinfo{pages}{e366}.
\newblock


\bibitem[Trinh and Liu(2021)]%
        {trinh2021examination}
\bibfield{author}{\bibinfo{person}{Loc Trinh} {and} \bibinfo{person}{Yan Liu}.} \bibinfo{year}{2021}\natexlab{}.
\newblock \bibinfo{title}{An examination of fairness of AI models for deepfake detection}.
\newblock
\newblock
\showeprint[arxiv]{2105.00558}~[cs.CV]


\bibitem[Walker and Sleath(2017)]%
        {walker2017systematic}
\bibfield{author}{\bibinfo{person}{Kate Walker} {and} \bibinfo{person}{Emma Sleath}.} \bibinfo{year}{2017}\natexlab{}.
\newblock \showarticletitle{A systematic review of the current knowledge regarding revenge pornography and non-consensual sharing of sexually explicit media}.
\newblock \bibinfo{journal}{\emph{Aggression and Violent Behavior}}  \bibinfo{volume}{36} (\bibinfo{year}{2017}), \bibinfo{pages}{9--24}.
\newblock
\urldef\tempurl%
\url{https://doi.org/10.1016/j.avb.2017.06.010}
\showDOI{\tempurl}


\bibitem[Wang(2022)]%
        {wang2022don}
\bibfield{author}{\bibinfo{person}{Moncarol~Y Wang}.} \bibinfo{year}{2022}\natexlab{}.
\newblock \showarticletitle{Don't believe your eyes: Fighting deepfaked nonconsensual pornography with tort law}.
\newblock \bibinfo{journal}{\emph{University of Chicago Legal Forum}} (\bibinfo{year}{2022}), \bibinfo{pages}{415--445}.
\newblock


\bibitem[Westerlund(2019)]%
        {westerlund2019emergence}
\bibfield{author}{\bibinfo{person}{Mika Westerlund}.} \bibinfo{year}{2019}\natexlab{}.
\newblock \showarticletitle{The emergence of deepfake technology: A review}.
\newblock \bibinfo{journal}{\emph{Technology Innovation Management Review}} \bibinfo{volume}{9}, \bibinfo{number}{11} (\bibinfo{year}{2019}), \bibinfo{pages}{40--53}.
\newblock
\urldef\tempurl%
\url{https://doi.org/10.22215/timreview/1282}
\showDOI{\tempurl}


\bibitem[Widder et~al\mbox{.}(2022)]%
        {widder2022limits}
\bibfield{author}{\bibinfo{person}{David~Gray Widder}, \bibinfo{person}{Dawn Nafus}, \bibinfo{person}{Laura Dabbish}, {and} \bibinfo{person}{James Herbsleb}.} \bibinfo{year}{2022}\natexlab{}.
\newblock \showarticletitle{Limits and possibilities for “Ethical AI” in open source: A study of deepfakes}. In \bibinfo{booktitle}{\emph{Proceedings of the 2022 ACM Conference on Fairness, Accountability, and Transparency}}. \bibinfo{pages}{2035--2046}.
\newblock
\urldef\tempurl%
\url{https://doi.org/10.1080/23742917.2023.2192888}
\showURL{%
\tempurl}


\bibitem[Wodajo and Atnafu(2021)]%
        {wodajo2021deepfake}
\bibfield{author}{\bibinfo{person}{Deressa Wodajo} {and} \bibinfo{person}{Solomon Atnafu}.} \bibinfo{year}{2021}\natexlab{}.
\newblock \showarticletitle{Deepfake video detection using convolutional vision transformer}.
\newblock \bibinfo{journal}{\emph{arXiv preprint arXiv:2102.11126}} (\bibinfo{year}{2021}).
\newblock


\bibitem[Xu et~al\mbox{.}(2022)]%
        {xu2022comprehensive}
\bibfield{author}{\bibinfo{person}{Ying Xu}, \bibinfo{person}{Philipp Terh{\"o}rst}, \bibinfo{person}{Kiran Raja}, {and} \bibinfo{person}{Marius Pedersen}.} \bibinfo{year}{2022}\natexlab{}.
\newblock \showarticletitle{A comprehensive analysis of AI biases in deepfake detection with massively annotated databases}.
\newblock \bibinfo{journal}{\emph{arXiv preprint arXiv:2208.05845}} (\bibinfo{year}{2022}).
\newblock


\end{thebibliography}
%\newpage
\appendix
\begin{table*} [!htb]
  \renewcommand\thetable{A1} 
  \captionof{table}{Participant Demographics. (Gender may not sum to 100\% due to the small fraction of respondents who preferred to self-describe, not disclose, or reported as non-binary.)}
\centering
\begin{tabular}{lccccc}\hline
  \hline
  \textit{Country} & \textbf{Australia}&\textbf{Belgium}&\textbf{Denmark}& \textbf{France}&\textbf{Mexico}\\ 
  \hline
\textit{Language(s) survey was offered in} & English & French & Danish & French & Spanish\\
\textit&&Dutch&&\\
   \hline
  \textit{Respondent count} & 1651 & 1617&1639&1636&1714\\
  \hline
  \textit{Gender}& 51\% women & 50\% women & 51\% women & 52\% women & 51\% women\\
  & 48\% men & 48\% men & 49\% men & 46\% men & 48\% men\\
  \hline
  \textit{Age} 18-24&10\% & 9\% & 9\% & 9\% &16\%\\
\hspace{6mm}25-34 & 10\% & 9\% & 12\% & 8\% & 13\%\\
\hspace{6mm}35-44 & 11\% & 10\% & 5\% & 9\% & 12\%\\
\hspace{6mm}45-54 & 9\% & 10\% & 10\% & 10\% & 10\%\\
\hspace{6mm}55-64 & 9\% & 10\% & 8\% & 11\% & 8\%\\
\hspace{6mm}65+ & 12\% & 11\% & 15\% & 12\% & 4\%\\
\hline
\textit{Sexual}  \hspace{10mm}Heterosexual & 87\% & 85\% & 91\% & 85\% & 84\% \\
\textit{orientation} \hspace{4mm} Gay\slash Lesbian & 4\% & 4\% & 3\% & 4\% & 3\%\\
\hspace{19mm} Bisexual & 5\% & 6\% & 3\% & 4\% & 5\%\\
\hspace{19mm} Prefer to self-describe & 2\% & 2\% &1\% &2\% &2\%\\
\hspace{19mm} Prefer not to say & 3\% & 3\% & 3\% &5\% &6\%\\
\hline
\hline
 \textit{Country} & \textbf{Netherlands}&\textbf{Poland}&\textbf{South Korea}& \textbf{Spain}&\textbf{USA}\\ 
  \hline
\textit{Languages} & Dutch & Polish & South Korean & Spanish & English\\
   \hline
  \textit{Respondent count} & 1665 & 1639&1641&1711&1780\\
  \hline
  \textit{Gender} & 51\% women & 52\% women & 50\% women & 52\% women & 51\% women\\
  & 49\% men & 47\% men & 50\% men & 48\% men & 47\% men\\
  \hline
  \textit{Age} 18-24&10\% & 9\% & 9\% & 8\% &11\%\\
\hspace{6mm}25-34 & 9\% & 10\% & 9\% & 8\% & 12\%\\
\hspace{6mm}35-44 & 9\% & 11\% & 11\% & 12\% & 10\%\\
\hspace{6mm}45-54 & 11\% & 9\% & 11\% & 12\% & 10\%\\
\hspace{6mm}55-64 & 9\% & 11\% & 13\% & 13\% & 10\%\\
\hspace{6mm}65+ & 14\% & 9\% & 5\% & 8\% & 11\%\\
\hline
\textit{Sexual}  \hspace{10mm}Heterosexual & 86\% & 86\% & 68\% & 89\% & 82\% \\
\textit{orientation} \hspace{4mm} Gay\slash Lesbian & 4\% & 2\% & 3\% & 3\% & 7\%\\
\hspace{19mm} Bisexual & 5\% & 3\% & 15\% & 4\% & 7\%\\
\hspace{19mm} Prefer to self-describe & 1\% & 2\% & 5\% & 1\% & 2\%\\
\hspace{20mm}Prefer not to say & 3\% & 7\%&10\% &3\% &4\%\\
\hline
\hline
\end{tabular}
\end{table*}

\begin{table*}
  \renewcommand\thetable{A2}
\centering
  \captionof{table}{Percentage of Respondents Indicating  Victimization or Perpetration Experience via Deepfake Pornography.} %\label{tab:title} 
\begin{tabular}{lrr}
  \hline
 \textbf{Country} & 
 \textbf{Victimization (95\% CI)} &\textbf{Perpetration (95\% CI)} \\ 
  \hline
 Australia & 3.7\% (2.8\%-4.7\%) & 2.4\%(1.6\%-3.3\%) \\ 
   Belgium & 0.7\% (0.4\%-1.2\%) &0.3\% (0.1\%-0.7\%)\\ 
   Denmark & 1.0\% (0.6\%-1.6\%) &0.7\% (0.4\%-1.2\%)\\ 
   France & 1.6\%(1.1\%-2.3\%)&1.1\% (0.7\%-1.8\%) \\ 
   Mexico & 2.9\% (2.2\%-3.8\%) &1.2\% (0.8\%-1.8\%)\\ 
   Netherlands & 1.0\% (0.6\%-1.6\%) &0.4\% (0.1\%-0.7\%)\\ 
  Poland & 0.9\% (0.5\%-1.4\%)&0.5\%(0.3\%-1.0\%)\\ 
  South Korea & 3.1\%(2.3\%-4.0\%) &1.2\% (0.7\%-1.8\%)\\ 
   Spain & 1.4\% (0.9\%-2.1\%) &0.8\% (0.4\%-1.3\%)\\ 
   USA & 2.3\% (1.6\%-3.1\%) & 2.6\% (1.8\%-3.5\%)\\ 
   \hline
\end{tabular}
\end{table*}

\begin{figure*}[!htb]
  \renewcommand\thefigure{A1}
\centering
\includegraphics[scale=.55]{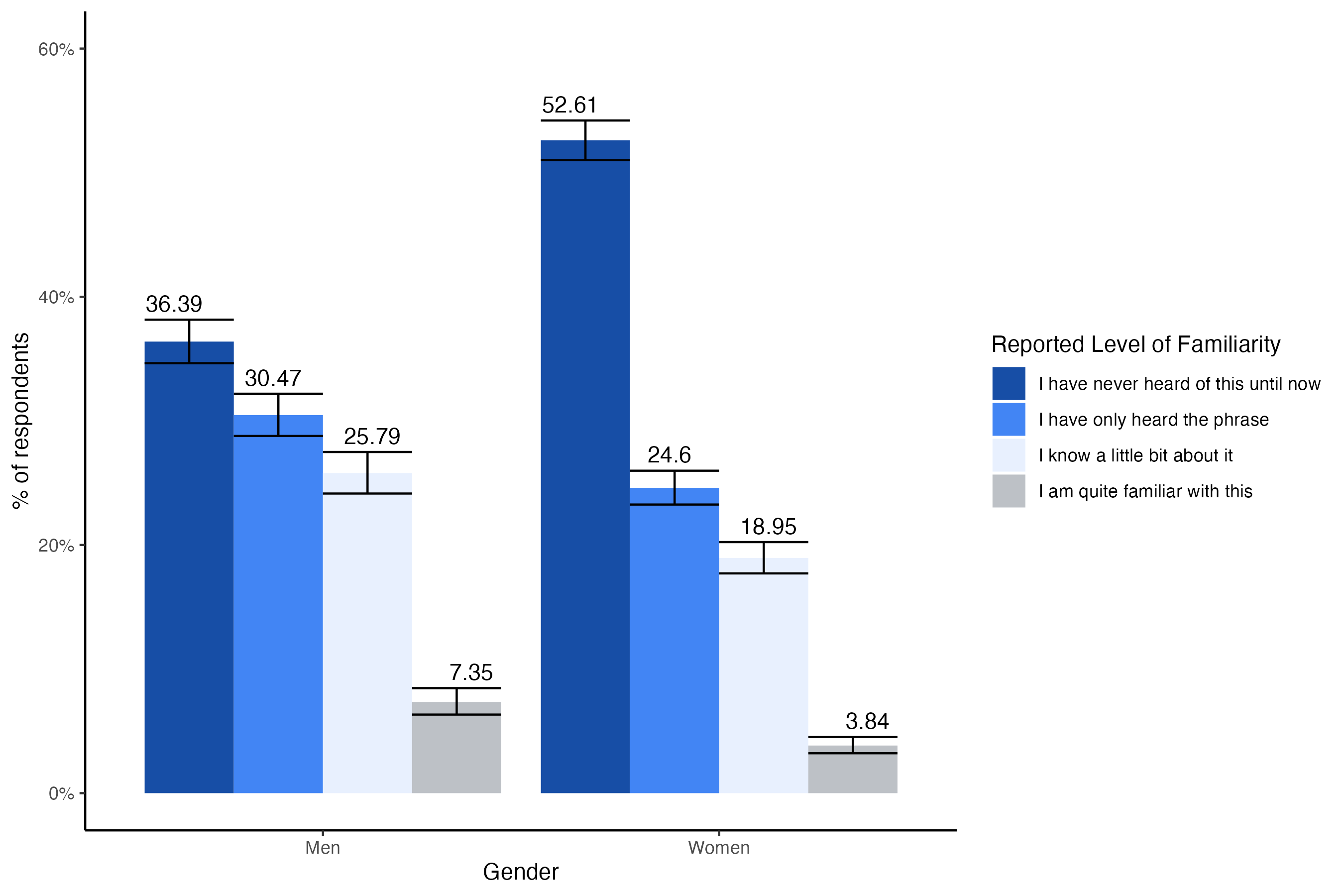}
\caption{Familiarity with Deepfake Pornography by Gender.}
\Description{A stacked barchart with gender on the x axis and \% of respondents endorsing each response option, which is a four point Likert scale from ``I have never heard of this until now'' to ``I am quite familiar with this.'' Men are more likely than women to fall into the ``I am quite familiar with this'' or ``I know a little bit about it'' categories (33\% compared to 23\%).}
\end{figure*}

\begin{figure*}[!htb]
  \renewcommand\thefigure{A2}

\centering
\includegraphics[scale=.7]{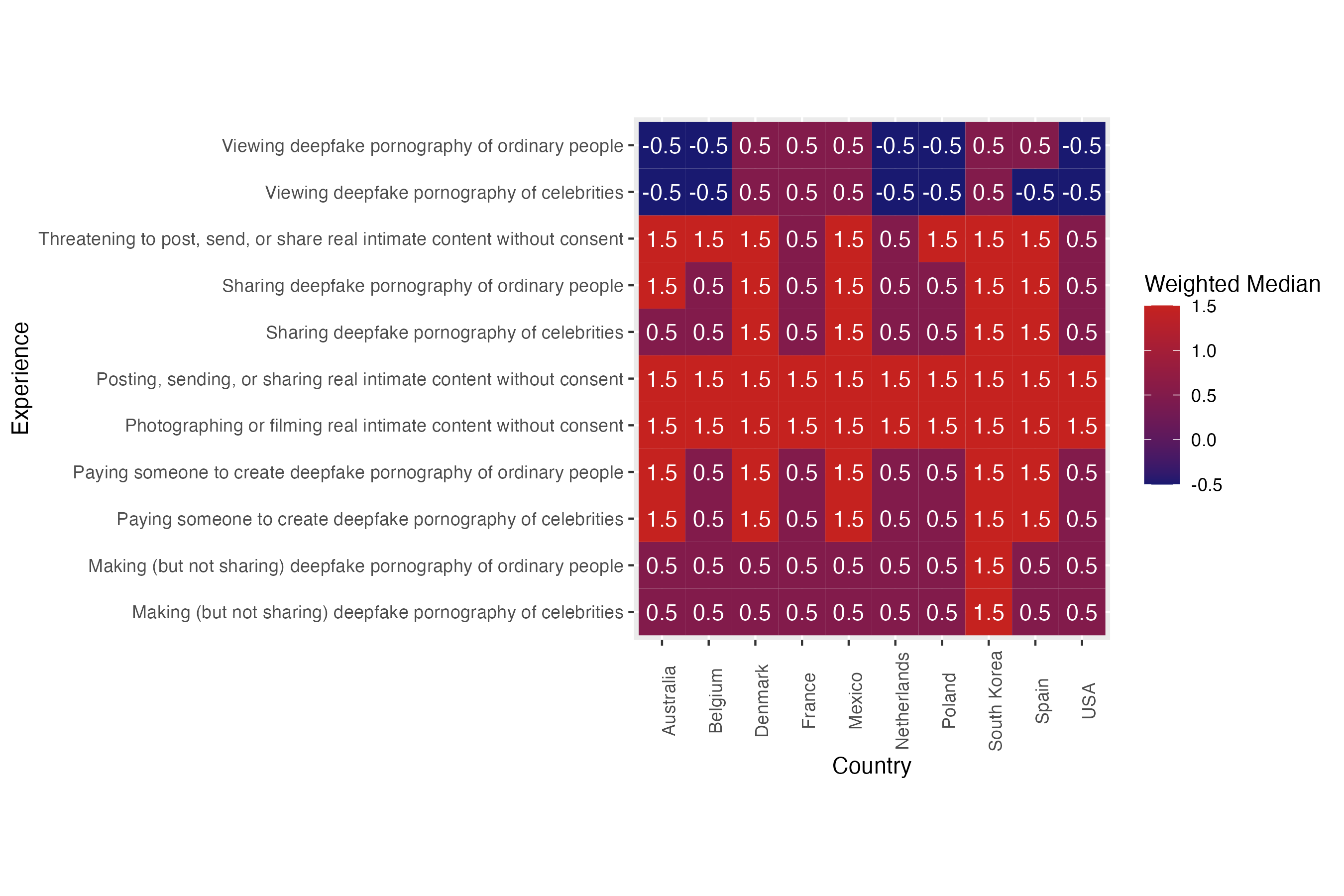}
\caption{Heatmap of Criminalization Attitudes by Country. Scale ranges from -2 (``Definitely should not be a crime'') to 2 (``Definitely should be a crime''). Brighter red is more deserving of criminalization, darker blue is less deserving of criminalization. Numbers shown are medians.}
\label{fig:label}
\Description[Heat map with country on the x axis and experience type on the y axis]{Heat map with country on the x axis and experience type on the y axis. Weighted medians are presented. The most deserving of criminalization behaviors are those that include real (as opposed to synthetic) material, and scores are typically 0.5 or 1.5, with the exception of the viewing of deepfake pornography, which is more commonly a -0.5.}
\end{figure*}

\section*{Appendix}
\begin{figure*}
  \renewcommand\thefigure{A3}

\centering
\includegraphics[scale=.65]{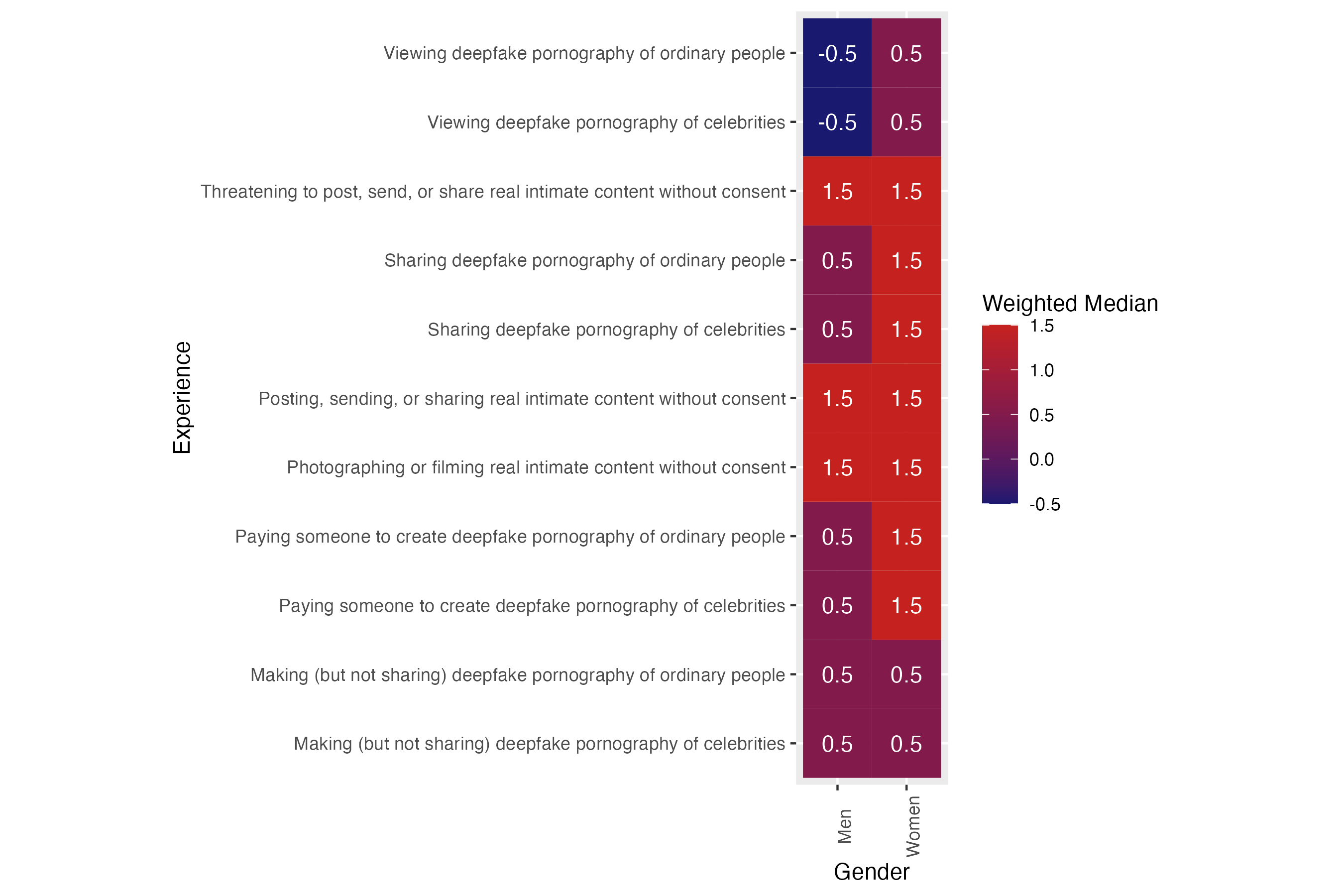}
\caption{Heatmap of Median Criminalization Attitudes by Gender. Scale ranges from -2 (``Definitely should not be a crime'') to 2 (``Definitely should be a crime''). Brighter red is more deserving of criminalization, darker blue is less deserving of criminalization. Numbers shown are medians.}
\label{fig:label}
\Description[Heat map with gender on the x axis and experience type on the y axis]{Heat map with country on the x axis and experience type on the y axis. Weighted medians are presented. Women and men tend to agree with regard to real content, while women consider the creating, sharing, and viewing of deepfake pornography to be more deserving of criminalization as compared to men}
\end{figure*}

\end{document}